\documentclass[preprint,12pt,3p]{elsarticle}

\usepackage{changepage}
\usepackage{lipsum}

\usepackage{array}
\usepackage{booktabs}
\usepackage{amssymb}
\usepackage{algorithm}
\renewcommand{\vec}[1]{\mathbf{#1}}

\usepackage{mathtools}
\usepackage{algpseudocode}
\usepackage{amsmath}
\usepackage{amssymb}
\usepackage{natbib}
\usepackage{verbatim}
\usepackage{hyperref} 
\usepackage{bm}

 \bibpunct[, ]{(}{)}{,}{a}{}{,}%
 \def\bibsep{\smallskipamount}%
\hypersetup{colorlinks=true}

\newcommand\blfootnote[1]{%
  \begingroup
  \renewcommand\thefootnote{}\footnote{#1}%
  \addtocounter{footnote}{-1}%
  \endgroup
}

\usepackage{textcomp}

\usepackage{lipsum}
\makeatletter
\def\ps@pprintTitle{%
 \let\@oddhead\@empty
 \let\@evenhead\@empty
 \def\@oddfoot{}%
 \let\@evenfoot\@oddfoot}
\makeatother
\begin{document}

\begin{frontmatter}

\title{A Framework to Integrate Mode Choice in the Design of Mobility-on-Demand Systems}
\blfootnote{Accepted manuscript for Transportation Research Part C, https://doi.org/10.1016/j.trc.2018.09.022}
\blfootnote{\textcopyright2018. This manuscript version is made available under the CC-BY-NC-ND 4.0 license http://creativecommons.org/licenses/by-nc-nd/4.0/}

\author[label1]{Yang Liu\corref{cor1}\fnref{myfootnote}}
\address[label1]{School of Civil and Environmental
Engineering, Cornell University, Ithaca, NY 14853}
\fntext[myfootnote]{These authors contributed equally to this work. }

\cortext[cor1]{Corresponding author}

\ead{yl2464@cornell.edu}

\author[label1]{Prateek Bansal\fnref{myfootnote}}
\ead{pb422@cornell.edu}

\author[label1]{Ricardo Daziano}
\ead{daziano@cornell.edu}

\author[label1]{Samitha Samaranayake}

\ead{samitha@cornell.edu}

\begin{abstract}
Mobility-on-Demand (MoD) systems are generally designed and analyzed for a fixed and exogenous demand, but such frameworks fail to answer questions about the impact of these services on the urban transportation system, such as the effect of induced demand and the implications for transit ridership. In this study, we propose a unified framework to design, optimize and analyze MoD operations within a multimodal transportation system where the demand for a travel mode is a function of its level of service. An application of Bayesian optimization (BO) to derive the optimal supply-side MoD parameters (e.g., fleet size and fare) is also illustrated. The proposed framework is calibrated using the taxi demand data in Manhattan, New York. Travel demand is served by public transit and MoD services of varying passenger capacities (1, 4 and 10), and passengers are predicted to choose travel modes according to a mode choice model. This choice model is estimated using stated preference data collected in New York City. The convergence of the multimodal supply-demand system and the superiority of the BO-based optimization method over earlier approaches are established through numerical experiments. We finally consider a policy intervention where the government imposes a tax on the ride-hailing service and illustrate how the proposed framework can quantify the pros and cons of such policies for different stakeholders. 
\end{abstract}


\end{frontmatter}


\section{Introduction}
\label{Sec: intro}
The rapid growth of Mobility-on-Demand (MoD) services such as Uber and Lyft is disrupting urban mobility. Over the last few years, this disruption has led to much interest in studying the design, management and impacts of these systems. Several analytical and simulation-based studies have focused on the efficient design of the MoD system -- allowing passengers to share rides and managing the fleet to serve travel demand with the minimum fleet size, passenger's waiting time, and vehicle miles traveled. In one of the first studies to develop a practical framework for managing a large-scale MoD fleet, \cite{ma2013t} devised a heuristic-based taxi dispatching strategy and fare management system that could handle operations of large urban scaled fleets. To incorporate ridepooling (capacity 2), \cite{santi2014quantifying} developed the notion of a shareability graph, and showed that almost all the taxi demand in Manhattan, New York (around 150 million annual trips as of 2011) could be served by pairing up requests (2 requests per taxi). Pushing this result further, \cite{alonso2017demand} presented a computationally efficient \emph{anytime optimal} algorithm that can enable real-time high capacity ridepooling (up to 10 riders per vehicle) at the scale of Manhattan. Results show that 98\% of the taxi demand in Manhattan can be served by 3000 vehicles of capacity 4 instead of the current fleet of 13586 active taxis. \par

With recent advancements in automation technologies and government regulations enabling this technology, autonomous MoD (AMoD) systems\footnote{Note that the design framework of MoD and AMoD is similar if behavioral aspects and driver economics are omitted.} are also gaining research interest \citep{levin2016multiclass, spieser2016shared}. For example, an agent-based simulation of AMoD services are conducted for both Austin, Texas \citep{fagnant2015operations} and Melbourne, Australia \citep{dia2017autonomous}. Both studies concluded that AMoD services can satisfy travel demand with around one-tenth of privately-owned vehicles. A few simulation-based studies also subsequently quantified the impact of MoD and AMoD services on the environment \citep{fagnant2014travel}, congestion \citep{fiedler2017impact}, and parking \citep{zhang2015exploring}. \par

In areas where public transit serves a large proportion of demand, the interaction between MoD services and transit should not be overlooked. Considering MoD services as a complementary mode of transit, a few studies have designed a bimodal MoD system to solve the first and last mile problem \citep{vakayil2017integrating, ma2017demand, shen2017embedding, moorthy2017shared}. \cite{ma2017demand} developed a real-time dispatching policy for MOD services in cooperation with existing mass transit service and tested it on a real network between Luxembourg City and its French-side cross-border area. \cite{shen2017embedding} studied the integration of MoD service with Singapore transit using agent-based simulation. The authors concluded that the integrated system is financially viable and has potential to make transit attractive by reducing out-of-vehicle time. \par

However, all these studies assumed a fixed and exogenous demand for MoD services. Most models assume a waiting time threshold with the demand that cannot be picked up within that threshold being considered \emph{unsatisfied} or \emph{drop-out} demand. The inherent assumption here is that passengers (from a certain demand set) that can be served by MoD will necessarily choose it. In practice, this may not be true: MoD services co-exist with other travel modes (e.g., transit and walking). The demand for MoD is not only a function of its attributes (e.g., price, travel time), but also depends on characteristics of the competing travel modes as well based on individual preferences. Therefore, even if a passenger can be served by MoD services, they may choose other travel modes for a number of reasons. \par

Moreover, optimizing the operational performance of AMoD or MoD systems under an exogenous demand cannot answer policy questions related to the extended impact of these services on the urban transportation system as a whole. Such questions include the impact of MoD services on transit ridership \citep{araldoimplementation}, vehicle ownership \citep{hao2017analysis}, and demand for parking \citep{zhang2017parking}. Only a handful of studies have considered an endogenous demand model where MoD or AMoD systems compete with other travel modes based on service quality. \cite{horl2016simulation} and \cite{araldoimplementation} integrated such models into the MATSim and SimMobility platforms which are multi-modal agent-based activity simulators. However, both simulators suffer from a common problem of relying on the scenario-based supply-demand analysis of virtual cities, not optimizing the supply-side parameters, and using a synthetic mode choice model. \par 

A few recent studies have considered supply-demand interactions in a multimodal transportation system in the presence of on-demand mobility service. These studies have done so in the context of agent-based stochastic user equilibrium with a day-to-day adjustment process~\citep{djavadian2017agenta}. In particular, \cite{djavadian2017agentb} models the dynamic adjustment process of an MoD operator. However, unlike our design-focused approach, these studies emphasize evaluating the sensitivity of different MoD operating policies as opposed to determining the optimal supply-side response. \par

The contribution of this study is twofold. The first contribution is to develop a unified framework which integrates mode choice models with a state-of-the-art system for modeling real-time on-demand mobility services with varying passenger capacities (based on \cite{alonso2017demand}). We illustrate the proposed framework by developing a multi-scale MoD fleet management system to serve the taxi demand in Manhattan where passengers are utility maximizers and choose a travel mode from a set of four mutually exclusive alternatives: ride-hailing service (capacity 1), ridepooling service (capacity 4), micro-transit service (capacity 10) and public transit (e.g. subway). The first three travel modes are MoD services assumed to be run by a single MoD provider. Note that the underlying mode choice model is calibrated with stated preference survey data from New York City. \par

The second contribution of this study is to implement and integrate a Bayesian Optimization (BO) based solver to find the optimal supply-side MoD parameters (e.g., fleet size, fare, etc). Historically, two types of methods have been used to determine the supply-side parameters of MoD systems: i) Heuristic methods to simultaneously optimize supply-side parameters and the vehicle routing problem (VRP) \citep{liu2009effective, repoussis2010solving}. Due to the complexity of VRP, these methods are inefficient for optimizing large-scale MoD operations; ii) Predefine candidate combinations of parameters (or scenarios), and then conduct a grid search to find the one that gives the optimal objective function value in the simulation \citep{li2010optimizing, brownell2014driverless, marczuk2015autonomous, azevedo2016microsimulation, boesch2016autonomous}. The performance of this approach highly depends on the quality of the predefined set of candidate solutions. In addition, grid search struggles in finding even near-optimal solutions in a high dimensional space in reasonably short time. In this article, we decouple the optimization of supply-side parameters and the MoD vehicle assignment problem. We consider the transportation simulation system as a black-box function, and use BO as a sequential search strategy to optimize the supply-side parameters. \par

In summary, the proposed framework would not only help in better understanding the influence of MoD services on urban mobility and answer system-level policy questions, but would also illustrate a method to tune the supply-side parameters that influence the demand. \par

We note that our current study considers the case of a single MoD operator and does not address competition between providers, which is a limitation of the current model. Realistically, several MoD and taxi services co-exist and their behaviors are dependent on competitive factors. Competition between these services and the resulting market equilibrium~\citep{wang2016pricing, qian2017taxi,heilker2018duopoly} and their impact on operator efficiency~\citep{sejourne2018price} have been studied independently. Integrating these competitive models in the proposed framework is a direction for future research. \par

The rest of this article is structured as follows. In Section~\ref{sec: formulation}, we formulate the mathematical problem and propose the unified framework. In particular, we describe two essential components: the mode choice model and the MoD simulation system. In Section~\ref{sec: solve}, we introduce a BO-based approach to find the optimal supply-side parameters. Section~\ref{sec: experiments} consists of numerical experiments on the Manhattan network, which are focused on the numerical convergence of the proposed framework, efficiency of the BO approach introduced in this work, and practical insights that the framework provides. Finally, Section~\ref{sec: conclusion} provides closing remarks and discusses possible directions for future research. \par

\section{Framework: Modeling Mobility-on-Demand Systems within a Multimodal Transportation System}\label{sec: formulation}
\begin{figure}[htb]
\centering
\includegraphics[width=1.05\textwidth]{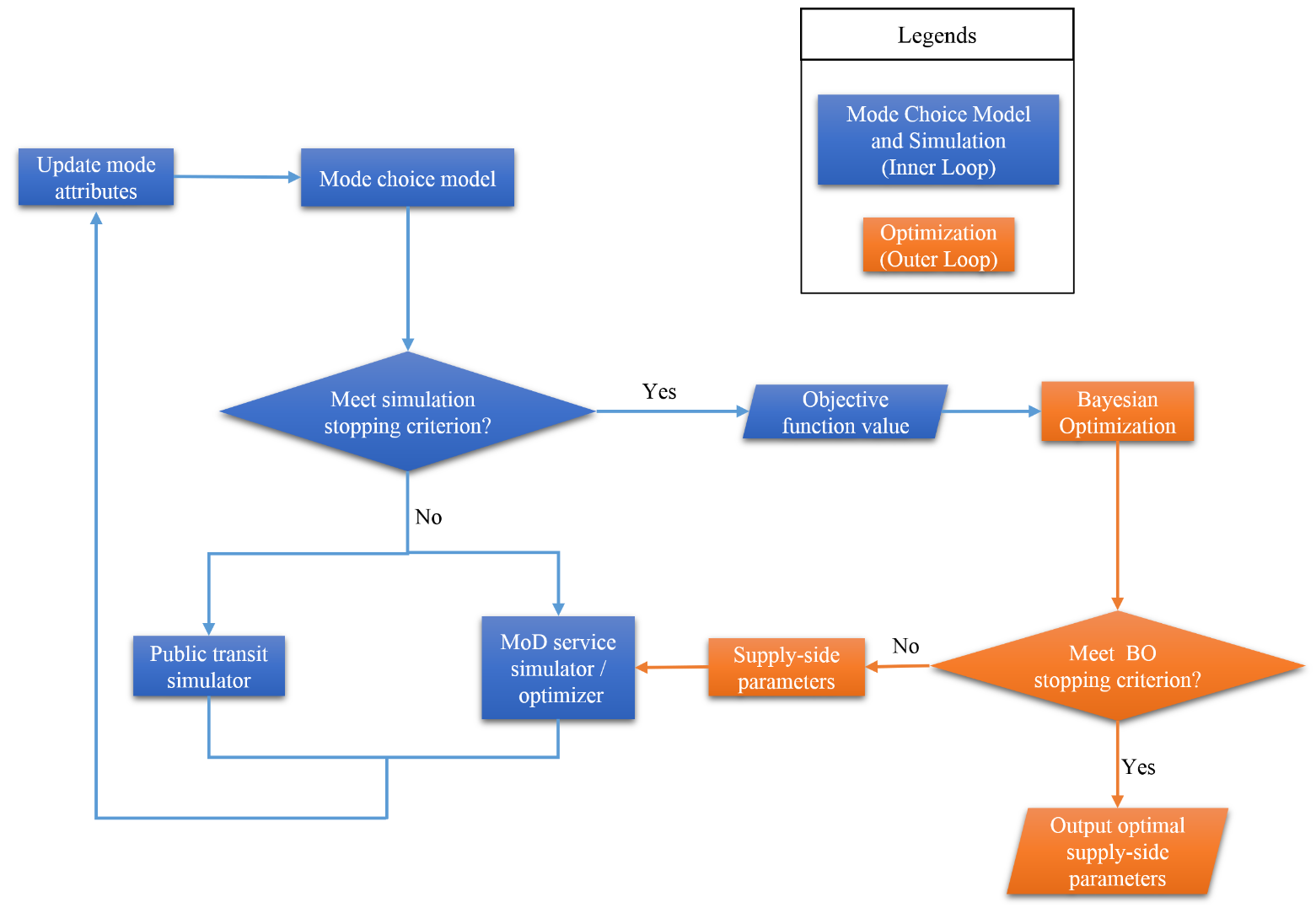}
\caption{\label{fig:procedure} A framework to optimize the supply-side parameters for MoD system with the integration of the mode choice model.}
\end{figure}

We consider a general transportation system that includes public transit and three classes of MoD services with varying passenger capacities (i.e., maximum occupancy). The public transit option is considered to be a complimentary travel mode to the MoD services, but note that the framework is not limited to this specific case\footnote{In the framework, the interplay between all travel modes is considered in the mode choice model. After the demand for each travel mode is generated, the simulator for each mode is run independently. Therefore, we can use any set of candidate travel modes for any purposes. }. The goal is to develop a simulation optimization framework to optimize both i) the system-level objective functions (e.g., MoD system operator's profit or consumer surplus) when the demand for MoD services is not exogenous and rather depends on their and other competing travel modes' level of service (e.g., waiting time, travel time, and travel cost), and ii) the transportation system operations given a certain set of supply and demand side parameters. The proposed simulation and optimization framework is shown in the Figure~\ref{fig:procedure}. \par

The proposed framework can be disentangled into inner and outer loops. The outer loop iteratively optimizes the supply-side parameters using BO and terminates when a stopping criterion of BO is satisfied. For a given set of supply-side parameters, the objective function (e.g., operator's profit) of the outer loop is evaluated at the equilibrium of the inner loop. The inner loop iteratively performs the following steps: i) evaluate mode choice probabilities, ii) simulate the demand for and operations of public transit and MoD services, iii) update mode-specific attributes (e.g., waiting time) as the inputs to the choice model in the next iteration, and iv) repeat until an equilibrium is reached, i.e. when the average difference in mode shares of two consecutive iterations is lower than a predefined threshold (see Section~\ref{sec: equilibrium}). In other words, if each iteration of the inner loop is interpreted as a ``day'', for a given set of supply-side parameters, passengers learn from the experienced historical level-of-service of travel modes so that they can make more informed mode choices on a given day. The inner loop terminates when passengers saturate their learning and make nearly consistent mode choices on consecutive days. \par

This section discusses the different components of the inner loop such as the mode choice model, and operations of public transit and MoD services. Section~\ref{sec: solve} discusses a BO-based approach to find the optimal supply-side parameters (outer loop). \par

\subsection{Mode Choice Model and Data}
To understand preferences of New Yorkers for MoD services, we conducted a stated-preference (SP) study. In an online survey, the respondents were asked about sociodemographics, travel characteristics, and various other opinions. The survey also had a discrete choice experiment (DCE) in which each respondent was asked to choose the best and the worst travel mode from a set of three choices: Uber (without ridesharing), UberPool\footnote{\emph{UberPool} represents MoD services with a passenger capacity of more than one.} (with ridesharing), and their current travel mode (the one used most often on their most frequent trips). Table \ref{tab:attribute} shows the attribute levels of the DCE design. In the case of monthly payment of trip and parking costs, a per trip cost was computed by dividing monthly cost with trip frequency. The per mile cost of a private car (\$.45) was obtained by summing the insurance, maintenance, and fuel cost (\href{https://newsroom.aaa.com/2016/04/driving-costs-hit-six-year-low-finds-aaa-2/}{AAA News Room, 2016}). In the DCE design, we made sure to constrain in-vehicle travel time (IVTT) and per mile cost of Uber to be less and more, respectively, than those of UberPool. Respondents were provided textual and visual information about all alternatives at the starting of DCE.\par

\begin{table}[h!]
		\centering
		\caption{Experiment Design for Mode Choices. \vspace{-.5cm}}
		\label{tab:attribute}
		\begin{adjustwidth}{0cm}{}
		\resizebox{1\textwidth}{!}{
		\begin{tabular}{lccc}
\hline
 & \textbf{Uber (Without Ridesharing)} & \textbf{UberPool (With Ridesharing)} & \textbf{Current Mode} \\ \hline
Walking and Waiting Time & 25\%, 50\%, 75\%, 100\% & 25\%, 50\%, 75\%, 100\% & asked (100\%) \\
In-vehicle Travel Time & 80\%, 95\%, 110\%, 125\% & 90\%, 105\%, 120\%,135\% & asked (100\%) \\
\begin{tabular}[c]{@{}l@{}}Trip Cost Per Mile (\$)\\ (Excluding Parking Cost)\end{tabular} & \begin{tabular}[c]{@{}c@{}}$0.55$, $0.70$, $0.85$, $1.0$, $1.2$\\ \end{tabular} & \begin{tabular}[c]{@{}c@{}}$0.45$, $0.60$, $0.70$, $0.80$\\ \end{tabular} & asked or computed \\
Parking Cost & 0 & 0 & asked \\
Powertrain & Gas, Electric & Gas, Electric & Gas \\
Automation & Yes, No & Yes, No & No \\ \hline
\multicolumn{4}{l} {\textbf{Note:} All \% are relative to the reference alternative. } \\
\end{tabular}}
\end{adjustwidth}
\end{table}	

To obtain priors for pivot-efficient DCE designs\footnote{In pivot-efficient designs, attribute levels shown to the respondents are pivoted from reference alternatives for each respondent. In this study, the travel mode used on the most frequent trips was considered as the reference alternative.}, we first conducted a pretest. A D-efficient design with zero priors containing 4 blocks (6 choice situations per block) was generated with the Ngene software \citep{choicemetrics2014ngene} for the pilot study. The attribute levels of Table \ref{tab:attribute} were used in the design. The online pilot survey was created using the web-based Qualtrics platform, but the survey was distributed among a continuous panel provided by Survey Sampling International (SSI, a professional survey firm) in February 2017. Those who drive for any MoD service or are younger than 18 years were considered ineligible to participate in the survey. Those who completed the survey in less than 10 minutes or provided conflicting responses (e.g., reported more children than household size) were discarded. After eliminating such responses, 298 (out of 397) completed responses were used as valid pretest observations for further discrete choice analysis. \par

We used prior parameter estimates from the pilot study to create a pivot-efficient design with 6 blocks (7 choice situations per block). All the attributes and attribute levels remained the same (Table \ref{tab:attribute}). Table \ref{tab:EXP} shows an example of the choice situation of the final mode choice experiment. \par

\begin{table}[h!]
		\centering
		\caption{An Instance of the Choice Experiment  \vspace{-.5cm}}
		\label{tab:EXP}
		\begin{adjustwidth}{0cm}{}
		\resizebox{1\textwidth}{!}{
		\begin{tabular}{lccc}
\hline
 & \textbf{Uber (Without Ridesharing)} & \textbf{UberPool (With Ridesharing)} & \textbf{Current Mode: Car} \\ \hline
Walking and Waiting time & 6 minutes & 9 minutes & 12 minutes \\
In-vehicle Travel Time & 38 minutes & 50 minutes & 48 minutes \\
\begin{tabular}[c]{@{}l@{}}Trip Cost\\ (Excluding Parking Cost)\end{tabular} & \$11 & \$8 & \$6 \\
Parking Cost & -- & -- & \$6 \\
Powertrain & Electric & Gas & Gas \\
Automation & Service with Driver & Automated (No Driver) & -- \\ \hline
\end{tabular}}
\end{adjustwidth}
	\end{table}	
	
We conducted the main study during October-November 2017. After data validation tests, preferences of 1507 (out of 1689) respondents were used in the model estimation. We estimated a multinomial logit (MNL) model and used MNL attribute valuation for prediction of mode choice probabilities, which are needed in the simulation. The closed-form choice probability MNL expressions allow a seamless integration in the MoD simulation framework\footnote{We tried nested logit specifications with different nesting structures, but in all scenarios, one of the nests' elasticity estimates was above one. In other words, nest correlation was out of the accepted range $[0,1]$ for compatibility with random utility maximization. This implies preferences of the sample are not aligned with nested logit correlation.}. Table \ref{tab:MNL} summarizes MNL parameter estimates. By dividing marginal utilities of attributes with that of trip cost, willingness-to-pay (WTP) estimates are derived. By using these marginal rates of substitution, the model provides evidence that New Yorkers are willing to pay \$25.9 and \$18.6 to save an hour of OVTT and IVTT, respectively. The recommended hourly value of travel time for local commute by passenger car in downstate New York is \$15.6 (\href{https://www.dot.ny.gov/divisions/engineering/design/dqab/hdm/hdm-repository/Recommended\%20Value\%20of\%20Time.pdf}{Department of Transportation, New York State, 2012}), which is close to our estimate of WTP to save an hour of IVTT. Another study of economic evaluation (\href{https://www.transportation.gov/sites/dot.dev/files/docs/vot_guidance_092811c.pdf}{US Department of Transportation, 2011}) estimates that walking, waiting, and transfer time in personal travel should be valued at \$19.10 - \$28.70 per hour. Our estimates of WTP to save an hour of OVTT falls in this range. Although WTP estimates can be transferred directly to the MoD simulation, alternative specific constants (ASCs) require recalibration (ASCs in SP studies are just manifestation of sample shares of alternatives). This process is consistent with the estimated mode choice model, but is not ideal because choice sets of SP study and the considered taxi demand are different\footnote{The recalibration of ASCs is much easier for the full travel demand model where all possible alternatives are available to travelers and their real mode shares are known.}. Details about recalibration of ASCs can be found in Sections~\ref{sec: ASC} and \ref{sec: scenario}. \par

\begin{table}[h!]
		\centering
		\caption{Multinomial logit model estimates \vspace{-.5cm}}
		\label{tab:MNL}
		\begin{adjustwidth}{1.5cm}{}
		\resizebox{.8\textwidth}{!}{
\begin{tabular}{lccc}
\hline
\textbf{Attributes} & \textbf{Coef.} & \textbf{Std. Err.} & \textbf{Z-stat} \\ \hline
Walking and waiting time (OVTT, in mins) & -0.032 & 0.0020 & -16.3 \\
In-vehicle travel time (IVTT, in mins) & -0.023 & 0.0025 & -9.3 \\
Trip cost & -0.074 & 0.0030 & -24.3 \\
Parking cost & -0.057 & 0.0091 & -6.3 \\
Electric & -0.041 & 0.0386 & -1.1 \\
Automation & -0.182 & 0.0409 & -4.5 \\ \hline
\textbf{ASC} & \textbf{} & \textbf{} & \textbf{} \\ \hline
Uber & -0.821 & 0.0646 & -12.7 \\
UberPool & -1.266 & 0.0519 & -24.4 \\
Carpool & 0.057 & 0.1594 & 0.4 \\
Ridesharing & -1.689 & 0.1592 & -10.6 \\
Transit & -0.232 & 0.0491 & -4.7 \\
Car & \multicolumn{3}{c}{base} \\ \hline
\textbf{Loglikelihood} & \multicolumn{3}{c}{-9762.281} \\
\textbf{Number of observations} & \multicolumn{3}{c}{1507} \\ \hline
\end{tabular}}
\end{adjustwidth}
	\end{table}

\subsection{Simulation of MoD and Transit Operations}
In this section, we introduce the system we use to simulate the operations of the MoD and transit services. \par

\subsubsection{Simulation of the MoD system}
We consider a fleet of vehicles with varying passenger capacities to satisfy the demand for MoD services. The operation of MoD services includes two main tasks: i) match travel requests to vehicles; ii) rebalance the idle vehicles to areas with potential future demand. Our formulation follows a state of the art framework which was recently proposed by \cite{alonso2017demand}. \par

The set of vehicles is denoted by $V = \{v_1, ... , v_n\}$. Each vehicle $v_i$ has a corresponding capacity $c_i$. The set of trip requests is denoted by $R = \{r_1, ..., r_n\}$. Multiple travelers are allowed to share one single ride. A \emph{passenger} is defined as a past request that has been picked up by a vehicle and is now en-route to its destination. For each request $r$, the waiting time is $w_r$, which is the difference between the pick-up time $t_r^p$ and the request time $t_r^r$. For each picked up request $r$, the total travel delay is defined as $\delta_r = t_r^d - t_r^*$, where $t_r^d$ is the drop-off time and $t_r^*$ is the earliest possible time at which the destination could be reached. $t_r^*$ is computed by following the shortest path between the origin $o_r$ and the destination $d_r$ with zero waiting time. We want to find the optimal assignment that minimizes a cost function $C$ under a set of constraints as follows: \par

\begin{itemize}
\item For each request $r$, the waiting time $w_r$ must be below the maximum waiting time $\Omega_r$. If one customer has waited for a time more than $\Omega_r$, the request is discarded in the simulation, and a large constant penalty will occur in the cost function. \par

\item For each picked up request $r$, the total delay $\delta_r$ must be lower than the maximum travel delay $\Delta_r$. \par

\item For each vehicle $v$ at any time, the number of passengers in the vehicle must not be larger than the capacity $c_v$ \par

\item The requests for travel via a specific MoD service type (e.g. ride-hailing service, ridepooling service or micro-transit service) can only be served by the vehicle fleet corresponding to the same service type. 
\end{itemize}

For convenience, the cost function $C$ is set to the sum of total delay and the penalty for unassigned requests in this work, but note that the objective function can be tuned for any other purposes. The total delay includes both the waiting time for all assigned requests and the in-vehicle delay caused by sharing with other passengers. \par

The MoD operation problem is essentially the same as the dynamic pickup and delivery problem \citep{savelsbergh1995general} with time windows, which has shown to be difficult to solve exactly for a large-scale demand \citep{nanry2000solving, ropke2006adaptive}. The anytime optimal algorithm in \cite{alonso2017demand} decouples the problem by first computing feasible trips from a pairwise shareability graph \citep{santi2014quantifying} and then finding the optimal trip assignment to vehicles by solving an Integer Linear Program (ILP) of reduced dimensionality. In our study, the assignment of the trip requests is processed at a frequency of every 60 seconds. Requests that have not been picked up in an assignment round remain in the request pool until the maximum waiting time constraint is violated. \par

Specifically, each round of the assignment simulation includes the following steps: \par

(1) Compute a pair-wise request-vehicle graph (RV-graph). RV-graph describes possible pairwise matching between vehicles and pick-up requests. In the graph, we connect two requests $r_i$ and $r_j$ if an empty virtual vehicle at the origin of one of them can serve both of them without violating any of the constraints; Similarly, we connect a request $r$ to a vehicle $v$ if $v$ can serve $r$ under the constraints. \par

(2) Construct the Request-Trip-Vehicle graph (RTV-graph) using the cliques of the RV-graph, which includes all the feasible trips and also the vehicles that can serve them. A feasible trip is defined as a set of requests that can be served by one vehicle under the constraints. In the RTV-graph, a request $r$ is connected to a trip $T$ if $T$ contains $r$; A trip $T$ is connected to a vehicle $v$ only if $v$ can complete $T$ without violating any constraints. The feasible trips and the edges in the graph are computed incrementally in trip length for each vehicle, which successfully reduces the search space \citep{alonso2017demand}. This step results in all the feasible assignments between the vehicles and the requests. \par

(3) Solve an ILP to compute the optimal assignment of vehicles to trips. Since each trip can possibly be served by multiple vehicles, the ILP needs to not only determine the assignment, but also ensure that each request and vehicle are assigned to only one trip. The ILP in \cite{alonso2017demand} is as follows: \par

\begin{equation}\label{old}
\begin{array}{ll@{}ll}
\text{minimize}  & \displaystyle\sum\limits_{i, j\in X_{TV}} c_{i, j} x_{i, j} + \sum\limits_{k \in \{1, ..., n\}}c_p \chi_k &\\
\text{subject to}& \displaystyle\sum\limits_{i \in \mathcal{I}_{V = j}^{T}}   x_{i, j} \leq 1  &\forall v_j \in V\\
                 & \sum\limits_{i\in \mathcal{I}_{R = k}^T} \sum\limits_{j\in \mathcal{I}_{T = i}^V} x_{i, j} + \chi_k = 1 &\forall r_k\in R
\end{array}
\end{equation}
where $X_{TV}$ is the set of all feasible assignments between trips and vehicles, $c_{i, j}$ is the cost of vehicle $j$ serving trip $i$, $c_{p}$ is the constant penalty for an unassigned request, which controls the trade-off between the total delay term and the penalty term. In our experiments, $c_{p}$ is set to a large constant to make sure that the objective is to minimize the total delay given that the maximum possible number of requests are served. If $c_p$ is set to a relatively small value, some requests may be rejected, since the cost of serving some requests may be higher than the penalty for not serving them. For example, if the pickup point is far away from all available vehicles. The decision variables are $x_{i, j}$ and $\chi_k$, where $x_{i, j} = 1$ if vehicle $j$ is assigned to trip $i$, $\chi_k = 1$ if the request $r_k$ is ignored in the assignment. In the constraints, there are three sets: the set of trips that can be served by vehicle $j$ is $\mathcal{I}_{V = j}^{T}$; the set of trips that contains request $r_k$ is $\mathcal{I}_{R = k}^T$; the set of vehicles that can serve trip $i$ is $\mathcal{I}_{T = i}^{V}$. The constraints ensure that: i) each vehicle will only serve one trip; ii) each request is either served by one vehicle or ignored ($\chi_k = 1$) in the assignment. In this article, we use the following formulation which reduces the number of variables by $|R|$ compared to the above formulation. \par

\begin{equation}\label{new}
\begin{array}{ll@{}ll}
\text{minimize}  & \displaystyle\sum\limits_{i, j\in X_{TV}} (c_{i, j} - c_{p} l_i)&\cdot x_{i, j}&\\
\text{subject to}& \displaystyle\sum\limits_{i \in \mathcal{I}_{V = j}^{T}}   x_{i, j} \leq 1,  &\forall v_j \in V\\&\sum\limits_{i\in \mathcal{I}_{R = k}^T} \sum\limits_{j\in \mathcal{I}_{T = i}^V} x_{i, j} \leq 1 \quad\quad &\forall r_k\in R\\
                 & x_{i, j} \in \{0,1\}, &\forall i, j\in X_{TV}
\end{array}
\end{equation}
where $l_i$ is the number of requests in trip $i$. In this formulation, we track $l_i$ for each trip $i$. A reward (negative penalty -$c_p \cdot l_i$) of serving all the requests in trip $i$ is added to the objective function. In formulation~\ref{old}, a penalty is imposed on the objective function for each unassigned request. In formulation~\ref{new}, we consider all the requests unassigned before the assignment, and thus a total penalty $c_p \cdot |R|$\footnote{This term is ignored in the objective function since adding constants does not affect the optimal solution. } is added to the objective function. Whenever we assign a vehicle $j$ to a trip $i$, the system gets a reward $c_p \cdot l_i$ for picking up $l_i$ requests in trip $i$. Therefore, the solution to formulation~\ref{new} is equivalent to the solutions to formulation~\ref{old} since the objective function is still the sum of the cost for each assigned trip plus the penalty of all unassigned requests, but the number of variables is reduced by $|R|$ because we do not need to include $\chi_k$ for each request $r_k \in R$ in the formulation. \par

(4) Rebalance the idle vehicles. After the assignment, there might still be idle vehicles and unsatisfied requests. It is reasonable to believe that more requests will be likely to appear in the neighborhood of the unsatisfied requests. Therefore, we solve a linear program as in \cite{alonso2017demand} to match the idle vehicles to the location of unsatisfied requests while minimizing the rebalance cost. \par

\subsubsection{Simulation of the Public Transit Service}
We assume that the network and operational characteristics of the existing public transit system (headway, travel time, etc) are known. To obtain the level of service attributes of the public transit in the mode choice model, we perform a transit network assignment for each origin-destination (OD). We first combine the road network we use in MoD service simulator and the public transit network. Then we use an all-pair shortest path algorithm to find the walk-transit-walk path for each OD pair. This procedure consists of the following steps: 

(1) Construct the public transit network. The public transit data is obtained from the General Transit Feed Specification (GTFS) dataset published by the transit authority through Google’s GTFS project. The public transit network topology is created according to the schedules and associated geographic information in the GTFS data. The weight on each link is the scheduled travel time between two transit stations. \par

(2) Combine the road network and the public transit network. The weight on the link between two road network nodes is set to the walking time ($\text{distance} / \text{walking speed}$). For each road network node, we add a link to connect it to each public transit network node within the walking range (e.g. 0.5 miles). As the origins and destinations are part of the nodes in the road network, the weight on the link connecting the road network node and the public transit node is set to a generalized cost, which is the sum of the walking time ($\text{distance} / \text{walking speed}$), the expected waiting time (half of the headway for the public transit line) and the trip fare (converted to seconds) for the public transit line. \par

(3) Compute the all-pairs shortest path using the combined network, and store them in a look-up table. During the simulation, we refer to the table to find the path and mode-specific attributes for public transit in the mode choice model. \par

\subsection{Updating Historical Trip Attributes and Inner Loop Stopping Criterion}
To model the memory and learning process of the travelers, we store and update the historical trip attributes for each OD after each iteration of the inner loop, and use these values as inputs to the mode choice model at the next iteration. Since the demand is sparsely distributed on the thousands of network nodes (e.g., 4092 nodes in our network), we use K-means clustering algorithm \citep{macqueen1967some} to spatially cluster the nodes. If the walking range for the passenger is $r$ miles, we determine the number of clusters $k$ by satisfying $\frac{\text{Total area}}{k} \approx 2\pi r^{2}$. The historical trip attributes are stored and updated at the cluster-level. \par

Due to the maximum waiting time $\Omega$ and the maximum delay $\Delta$ constraints, some of the demand for a certain MoD service may not be satisfied by the system. We assume that the demand that is unsatisfied (due to these constraints) will switch modes and take public transit. Therefore, for each cluster pair $(i, j)$, we also store the service rate $s_{i, j, m}$ for each MoD mode $m$, which is the proportion of travel demand served by mode $m$ for cluster pair $(i,j)$. \par

We compute the utility of the MoD services for the next iteration as follows (accounting for unsatisfied requests): Assume that for the cluster pair $(i, j)$, the utility of taking MoD mode $m$ and being picked up in the simulation is $U_{i, j, m}^{'}$, and the utility of taking transit is $U_{i, j, t}$. Then, the utility of taking the MoD mode $m$ in the next iteration $U_{i, j, m}$ is computed according to the following equation: \par

\begin{equation}
    U_{i, j, m} = s_{i, j, m} \cdot U_{i, j, m}^{'} + (1 - s_{i, j, m}) \cdot c_m \cdot U_{i, j, t}
\end{equation}

A constant penalty multiplier $c_m$ is used since transit was not the passenger's first choice in the last iteration. A high $c_m$ penalizes the utility function more if a request is not satisfied, and further decreases the corresponding MoD mode's share. In the numerical experiments, we use $c_m = 2$. This parameter should ideally be calibrated using a stated preference study, which we leave for the future research. $U_{i, j, m}^{'}$ and $U_{i, j, t}$ can be computed according to the historical trip attributes. After each iteration, the historical trip attributes are updated according to the following equation: \par

\begin{equation}
    \vec{H_{i, j, m}^{n + 1}} = \beta \vec{H_{i, j, m}^{n}} + (1 - \beta) \vec{I_{i, j, m}^{n}}
\end{equation}

where $\vec{H_{i, j, m}^{n}}$ is the set of historical trip attributes for cluster pair $(i, j)$ for mode $m$ at iteration $n$ which consists of the in-vehicle travel time, waiting time and the service rate, $\beta$ \footnote{In our numerical experiments, we let $\beta = 0.5$. } is a constant coefficient that controls the balances between the historical information and the new information, $\vec{I_{i, j, m}^{n}}$ is the trip attributes for cluster pair $(i, j)$ for mode $m$ which we obtain from the simulation in the current iteration by averaging the attributes for all ODs in the cluster pair. \par

The inner loop simulation procedure is iterated until the following stopping criterion is satisfied: \par

\begin{equation}\label{stop_cri}
Z^{n} = \frac{1}{|M|} \sum\limits_{m \in M} |S_m^{n} - S_m^{n - 1}|
\end{equation}
where $Z^{n}$ represents the average mode share difference among all the travel modes between iteration $n$ and $n - 1$, $M$ is the set of candidate travel modes, $S_{m}^{n}$ is the share of mode $m$ at iteration $n$. If $Z$ attains a value smaller than a predefined threshold of 0.01, the inner loop is terminated. In other words, the learning of travel demand saturates and system appears to achieve an equilibrium as a result of their consistent travel mode choices on consecutive "days". \par

\section{System Optimization using Bayesian Optimization} \label{sec: solve}
In Section~\ref{sec: formulation}, we discussed a simulation framework where the inner loop considers the supply-demand interaction for a given set of operational parameters of MoD services. In this section, we study the problem of optimizing these parameters which include fleet size and the pricing rules for MoD services with varying passenger capacities. \par 

The inner-loop simulation can be considered as a black-box function $f(\vec{x})$ for which the set of supply-side parameters $\vec{x}$ (i.e., decision variables) should be chosen such that the system performance metric at the equilibrium of inner loop attains its optimal value. The performance metric depends on the influence of different stakeholders. For example, whereas policy-makers might be interested in optimizing the consumer surplus and vehicle miles traveled, MoD operator might be interested in just maximizing the profit. \par

This optimization problem is difficult to solve because of the following properties: \par

\begin{itemize}
\item The objective function cannot be evaluated analytically and its derivatives are thus not easily available; \par

\item One can obtain observations of the objective function, but the function evaluation for even one set of decision variables is computationally expensive; \par

\item The function evaluation can be affected by simulation noise such as prediction of mode choice probability and other random factors (e.g. the initial location of the vehicles) in the MoD service simulator. \par
\end{itemize}

Whereas these characteristics make the problem intractable to solve using analytical optimization methods, Bayesian Optimization (BO) is an appropriate and powerful tool in such settings. BO is a sequential search strategy for the global optimization of an expensive black-box function $f(x)$ \citep{mockus2012bayesian}. It first emerged as a successful strategy in many machine learning applications \citep{bergstra2011algorithms, snoek2012practical, bergstra2013making, swersky2013multi, swersky2014raiders, yogatama2015bayesian}, and has lately been employed in many other areas including robotics \citep{lizotte2007automatic, calandra2016bayesian}, sensor networks \citep{garnett2010bayesian}, environmental
monitoring \citep{marchant2012bayesian}, information extraction and retrieval \citep{wang2014bayesian, li2018bayesian}, and game theory \citep{picheny2016bayesian}. 

Since the objective function does not have a closed-form expression in terms of the decision variables, BO treats it as a black-box function with some prior belief. Here \textit{prior} represents our belief about the space of possible objective functions. As we obtain more observations of the function, our belief about the objective function is updated by combining the prior and the likelihood of the data already acquired to get a potentially more informative posterior. The posterior distribution is then used to select the next set of decision variables for the function evaluation. \par

Specifically, a BO framework has two key ingredients: i) a probabilistic surrogate model for the expensive objective function $f$; ii) an acquisition function that is maximized to determines the next point for function evaluation \citep{shahriari2016taking}. The BO framework is demonstrated in Algorithm~\ref{Alg:BO}, where $S$ is the surrogate model, $\alpha$ is the acquisition function, and $\mathcal{D}_i = \{(\vec{x}_1, y_1), (\vec{x}_2, y_2), ..., (\vec{x}_i, y_i)\}$, which is the set of historical function observations until iteration $i$. Each observation is denoted as $(\vec{x}_i, y_i)$ where $\vec{x}_i$ is the set of inputs, and $y_i$ is the objective function value we obtain by evaluating the objective function $f$. \par

\begin{algorithm}[H]
\caption{Bayesian Optimization}
\begin{algorithmic}[1]
\For{$i = 1, 2, ...,$}
\State $\vec{x}_{i} = \operatorname*{argmax}\limits_\vec{x} \alpha(\vec{x}; \mathcal{D}_{i - 1})$
\State Evaluate the objective function using $\vec{x_i}$ to get $y_{i}$
\State $\mathcal{D}_{i} = \mathcal{D}_{i - 1} \cup (\vec{x}_{i}, y_{i})$
\State Update $S$.
\EndFor
\State Output the best $y$.
\end{algorithmic}\label{Alg:BO}
\end{algorithm}

We now provide a brief introduction of these two ingredients. Readers can refer to \cite{brochu2010tutorial} and \cite{shahriari2016taking} for more details of the method. \par

\subsection{Surrogate Model}
Since evaluating $f(x)$ is expensive, the BO framework builds a surrogate model using the historical observations to approximate the objective function. In this work, we adopt the Gaussian Process (GP) as the surrogate model, which is the most commonly-used model in BO \citep{mockus1994application}. GP is a stochastic process which can be seen as a collection of random variables where any finite set of random variables follows a multivariate Gaussian distribution. Assume that the set of points we use to build the GP is denoted by $\mathcal{X}$, then the GP is a fully specified by its mean function $m: \mathcal{X} \rightarrow \mathbb{R}$ and covariance function $k: \mathcal{X} \times \mathcal{X} \rightarrow \mathbb{R}$. Intuitively, GP is a distribution over functions. Given a point $\vec{x}$, it will return the mean and the variance of a normally distributed variable $y$ which is a prediction for $f(\vec{x})$. \par

For simplicity, the mean function $m(\cdot)$ is usually defined to be zero function, i.e., $m(\cdot) = \vec{0}$ \citep{brochu2010tutorial, marchant2012bayesian, wang2014bayesian}. Assume that we have historical observations $\mathcal{D}_{i}$, and we want to use GP to predict $f(\vec{x^{*}})$ on an arbitrary point $\vec{x}^{*}$. GP will return a normally distributed variable $y^{*}$ with both the mean and variance. Let $\vec{x}_{1:i} = [\vec{x}_1, \vec{x}_2, ..., \vec{x}_i]^{T}$ and $y_{1:i} = [y_1, y_2, ..., y_i]^{T}$. As the definition of GP reveals, the joint distribution of the historical observations and $y^{*}$ is as follows: \par

\begin{equation}
    \left[\begin{matrix}
       y_{1:i} \\
       y^{*}
      \end{matrix}
      \right] \sim \mathcal{N} \left(\vec{0}, \left[\begin{matrix}
      \vec{K} & \vec{k}_{\vec{x}^{*}} \\
      \vec{k}_{\vec{x}^{*}}^{T} & k(\vec{x}^{*}, \vec{x}^{*})
      \end{matrix}
      \right] \right)
\end{equation}
where $\vec{k}_{\vec{x}^{*}} = [k(\vec{x}^{*}, \vec{x}_1), k(\vec{x}^{*}, \vec{x}_{2}), ..., k(\vec{x}^{*}, \vec{x}_{i})]^T$, and $\vec{K}$ is the covariance matrix with each entry $\vec{K}_{(j, n)} = k(\vec{x}_j, \vec{x}_n)$. \par

Then, we can obtain the predictive distribution over $y^{*}$ as follows: \par

\begin{equation}
    y^{*}|\vec{x}_{1:i}, y_{1:i}, \vec{x}^{*} \sim \mathcal{N}(\mu(\vec{x}^{*}), \sigma(\vec{x}^{*}))
\end{equation}
where $\mu(\vec{x}^{*}) = \vec{k}_{\vec{x}^{*}}^{T}\vec{K}^{-1}y_{1:i}$, $\sigma(\vec{x}^{*}) = k(\vec{x}^{*}, \vec{x}^{*}) - \vec{k}_{\vec{x}^{*}}^{T}\vec{K}^{-1}\vec{k}_{\vec{x}^{*}}$. \par

The choice of covariance function $k$ for GP determines the smoothness properties of samples drawn from it \citep{brochu2010tutorial}. Similar to many other studies, we use the Matern kernel \citep{matern2013spatial}, which incorporates a smoothness parameter $\varsigma$ to provide flexibility in modeling functions:  
\begin{equation}
    k(\vec{x_{i}}, \vec{x_{j}}) = \frac{1}{2^{\varsigma - 1} \Gamma(\varsigma)} (2\sqrt{\varsigma} \|\vec{x_{i}} - \vec{x_{j}}\|)^\varsigma H_{\varsigma} (2\sqrt{\varsigma} \|\vec{x_{i}} - \vec{x_{j}}\|),
\end{equation}
where $\Gamma(\cdot)$ and $H_{\varsigma}(\cdot)$ are the Gamma function and the Bessel function of order $\varsigma$, respectively. When $\varsigma = 1/2$, the kernel reduces to the unsquared exponential kernel, and when $\varsigma \rightarrow \infty$, the kernel reduces to the squared exponential kernel. To provide appropriate smoothness, BO applications usually use $\varsigma = 3/2$ and $\varsigma = 5/2$ \citep{rasmussen2006gaussian}. In our experiments, we use $\varsigma = 5/2$. Readers can refer to \cite{hutter2013kernel} and \cite{ swersky2014raiders} for more details about the GP as well as its covariance functions.\par

\subsubsection{Acquisition Function}
The acquisition function is used to select the next realization of the decision variable for the evaluation of the objective function. It usually considers both the mean and variance of the predictions provided by the surrogate model. Intuitively, the process can be seen as maximizing the utility of the next sampling. \par

A good acquisition function is the one that finds an elegant trade-off between exploration and exploitation\footnote{Exploration represents searching in regions with high uncertainty but might have observations with high objective function values. Exploration helps in avoiding being trapped in a local optimum. Exploitation represents searching in the regions with high expected values of the objective function given the information provided by historical function evaluations.} while searching for the optimum of the objective function. The BO literature proposes many acquisition functions such as the probability of improvement (PI) \citep{kushner1964new}, the expected improvement (EI) \citep{szego1978towards}, and the upper confidence bound (UCB) \citep{ srinivas2009gaussian}. Similar to many other studies, we use UCB as the acquisition function. \par

\begin{equation}
    \text{UCB}(\vec{x}) = \mu(\vec{x}) + \kappa \sigma(\vec{x})
\end{equation}
where $\kappa$ is a hyperparameter which establishes a balance between the exploration and the exploitation. In our experiments, we employ a special case of UCB which is called GP-UCB \citep{srinivas2009gaussian, calandra2016bayesian} that casts the BO problem as a multi-armed bandit problem. In GP-UCB, $\kappa$ is automatilly updated according to the following euqation: \par

\begin{equation}
    \kappa = \sqrt{2 \log (\frac{n^{d/2 + 2}\pi^2}{3\delta})}
\end{equation}
where $n$ is the number of past objective function evaluations, $\delta \in (0, 1)$ is a parameter of choice, $d$ is the dimensionality of the search space. We use $\delta = 0.1$ the same as in \cite{srinivas2009gaussian}. \par

\subsection{Objective Function}
 In this article, we define the objective from the perspective of MoD service provider. Consider that all MoD operations are run by one service provider who is a profit maximizer. The profit is defined as the difference between the revenue and the operating cost. We use the ride-hailing service (capacity 1) as the base mode for the fare calculation of other MoD services and compute its trip fare according to the equation for UberX service \citep{Uber}. The fare for all other MoD modes $m$ with higher passenger capacities are computed by applying a discount factor $\gamma_m$ ($0 \leq \gamma_m \leq 1$) on the fare of the ride-hailing service. For a passenger $r$ that is served by the tier $m$ MoD service, the trip fare $f_r$ is computed as follows: \par
\begin{equation}\label{eqa: trip_cost}
    f_r = (1 - \gamma_m) \cdot \max (f_{min}, f_{base} + f_t \cdot t_r + f_d \cdot d_r)
\end{equation}
where $f_{min}$ and $f_{base}$ are the minimum fare and base fare for the ride-hailing service, $t_r$ and $d_r$ are the travel time and travel distance for $r$, $f_t$ is the time rate for the ride-hailing service, which is the fare per second, and $f_d$ is the distance rate for the ride-hailing service, which is the fare per mile. \par

The objective function $O(\bm{\gamma}, \vec{n})$ of the MoD service provider is computed as follows: \par

\begin{equation}
    O(\bm{\gamma}, \vec{n}) = \sum\limits_{i \in P} f_r - \sum\limits_{m \in M}\left[(c_{m, l} + s_m)\cdot n_m + c_{m, d} \cdot d_m\right]
\end{equation}
where $P$ is the set of all passengers, $c_{m, l}$ is the leasing cost for a tier $m$ MoD vehicle in the experiment time span, $s_m$ is the salary that the operator pays to tier $m$ driver\footnote{The drive salary can be ignored in the case of AMoD systems.}, $d_m$ is the total distance traveled by the tier $m$ MoD service and is obtained from the simulator, and $n_m$ and $c_{m, d}$ are the fleet size and operating cost (\$ per mile) of the tier $m$ MoD service. The set of discount factor and fleet size for each tier of MoD service are denoted as $\bm{\gamma}$ and $\vec{n}$, which are the decision variables. \par

Even though this work considers the objective from the perspective of a service provider, the proposed framework is more general and can be applied to designing policies, subsidies and regulatory strategies, guiding infrastructure planning and deployment, and operational management of transit services in the presence of MoD systems. \par

\section{Numerical Experiments} \label{sec: experiments}
In this section, we present a series of numerical experiments to show: i) the numerical convergence to an equilibrium of the mode choices within the multimodal system; ii) the calibration of the ASCs; iii) the performance of the BO-based optimization algorithm; iv) relevant applications of the proposed framework. In the following experiments, \textit{convergence to an equilibrium} represents that the system reaches a state where the average difference in mode shares of two consecutive iterations is lower than 1\%. The system is implemented using Python $3.5$, and all experiments are conducted on an Intel core I7 computer (3.4 gigahertz, 16 gigabytes RAM). \par

As a proxy for the real OD demands, we use the publicly available dataset of taxi trips in Manhattan, New York \citep{donovan2015using}. In our system, we serve this demand via either i) the ride-hailing service (capacity 1); ii) the ridepooling service (capacity 4); iii) the micro-transit service (capacity 10), and iv) public transit (e.g. subway). The first three modes are MoD services offered by one MoD service provider. We consider the pickup time for the taxi trips in the dataset to be their departure time . The network we use is the entire road network of Manhattan (4092 nodes and 9453 edges) \citep{santi2014quantifying, alonso2017demand}. The link travel time is given by the daily mean travel time, which is computed using the method in \cite{santi2014quantifying}. For public transit, we use the subway network provided by the Metropolitan Transportation Authority \citep{MTA}. The following experiments are conducted with respect to the rush hour demand (8 am to 9 am) on an arbitrary day (Monday the 6th of May, 2013). \par

The system has 5 decision variables: the fleet size of the ride-hailing service ($n_1$) and the fleet sizes and discount factors for the ridepooling and micro-transit services ($n_4$, $n_{10}$, $\gamma_4$, and $\gamma_{10}$). Note that at the initialization stage, there is no information about the OD-specific MoD attributes (e.g., waiting time and travel time). Therefore, we initialize the attributes as follows: Let $\Omega$ be the maximum waiting time allowed in the MoD system and $t_{o,d}$ be the shortest path travel time for each $(o, d)$. The travel time of the ride-hailing service, ridepooling service, and micro-transit service for $(o, d)$ is set to $t_{o,d}$, $1.2 t_{o,d}$, $1.5 t_{o,d}$ respectively and the waiting time is set to $0.3 \Omega$, $0.36 \Omega$, $0.45 \Omega$ respectively. In our experiments, we set $\Omega$ at 10 minutes. \par

To account for the labor cost of the drivers, in this study, we consider a driver salary of \$17 per hour for all MoD services, which is close to the hourly mean wage of taxi drivers in New York state (\href{https://www.bls.gov/oes/current/oes533041.htm}{United States Department of Labor, May 2017}). The leasing cost includes insurance, depreciation, maintenance, and other registration charges, and the operating cost includes fuel and tier cost. We compute these costs for different MoD services using statistics provided by \href{https://newsroom.aaa.com/2016/04/driving-costs-hit-six-year-low-finds-aaa-2/}{AAA News Room, 2016}. The leasing cost of \$11.97 per day per vehicle is obtained for the ride-hailing and ridepooling fleet of Sedan cars and \$19.32 per day per vehicle for the micro-transit fleet of Minivans. We normalize these numbers based on the proportion of daily demand (5.94\%) served during the experiment hour. The operating cost turns out to be \$0.1473 per mile for all MoD services.

\subsection{Convergence to an Equilibrium}\label{sec: equilibrium}
In this section, we illustrate the proposed multimodal system converging to an equilibrium with respect to mode choices using two test cases. In the test cases, we are not optimizing the system, but rather using a fixed set of supply-side parameters to test for the numerical convergence of the inner loop simulation to an equilibrium. The experiments are run for 20 iterations for each set of parameters. \par

\begin{figure}
\begin{minipage}[t]{0.5\linewidth}
\centering
\includegraphics[width=3.6in]{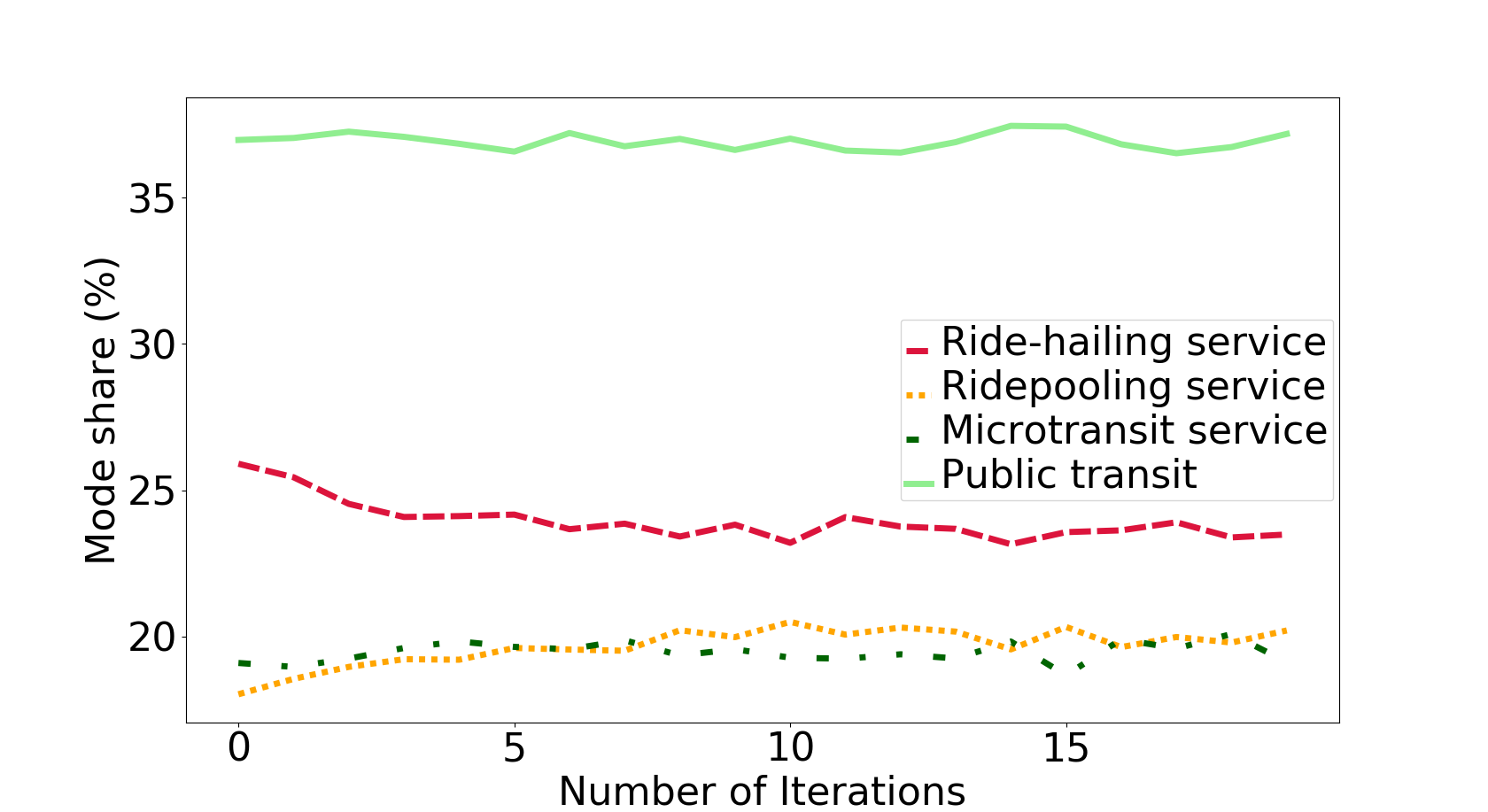}
\caption{\label{fig:test_case1} The mode share for the test case 1.}
\end{minipage}%
\begin{minipage}[t]{0.5\linewidth}
\centering
\includegraphics[width=3.6in]{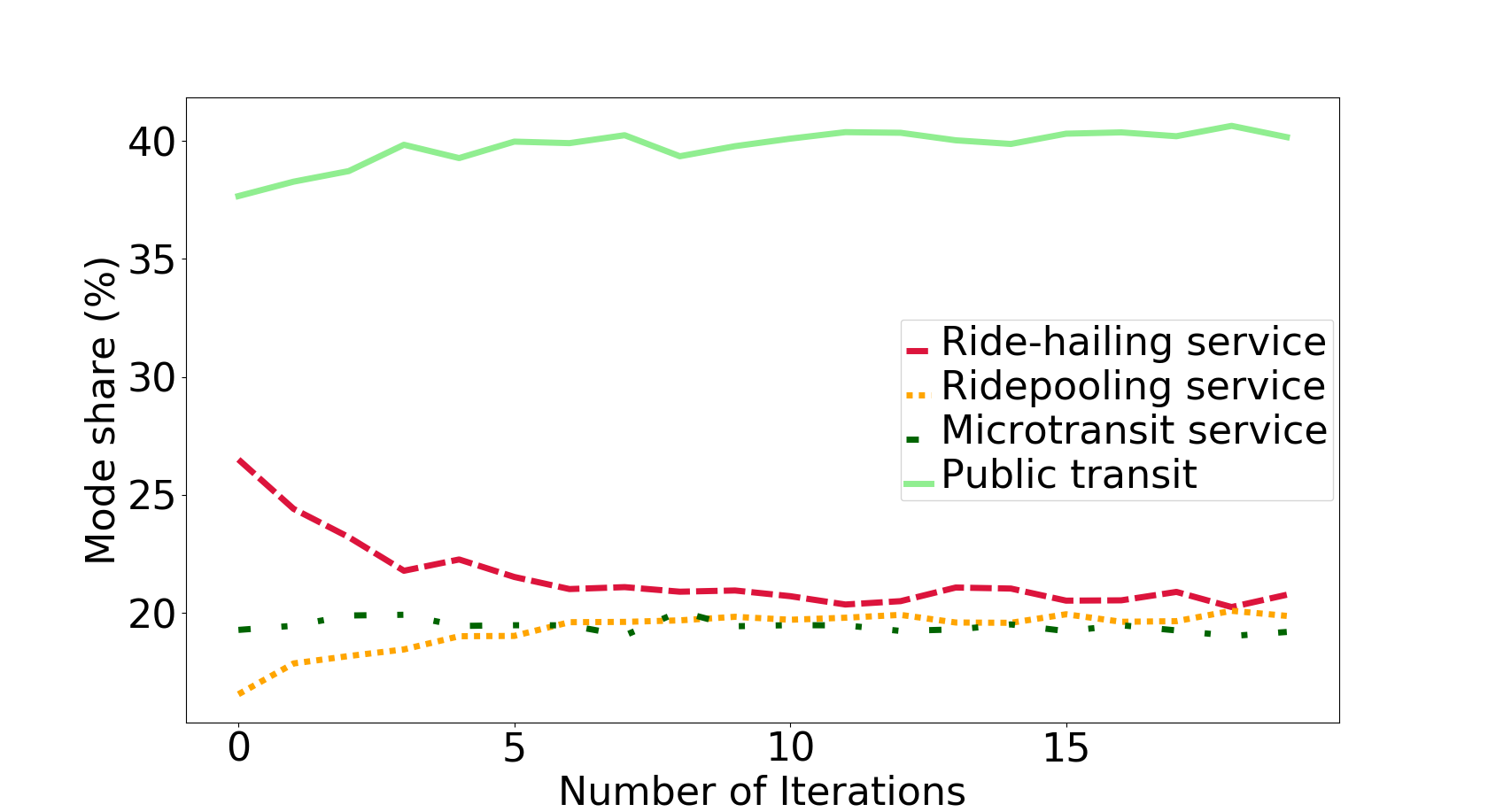}
\caption{The mode share for the test case 2.}
\label{fig:test_case2}
\end{minipage}
\end{figure}

The supply-side parameters for the first case are as follows: $n_1 = 800$, $n_4 = 1000$, $n_{10} = 500$, $\gamma_4 = 0.2$ and $\gamma_{10} = 0.4$. The iterative variation in the share of each travel mode is shown in Figure~\ref{fig:test_case1}. The share of each mode is relatively stable after around 5 iterations, which shows the numerical convergence of the system to an equilibrium. Note that the mode share is likely to fluctuate a little even after reaching equilibrium. This is just a manifestation of the simulation noise of the probabilistic mode choice model. In this test case, the mode share at equilibrium is not significantly different from the mode share at the beginning because the initial waiting time and travel time are close to the ones that the MoD services provider can offer given the supply-side parameters. \par

We set $n_1 = 100$, $n_4 = 1000$, $n_{10} = 50$, $\gamma_4 = 0.05$ and $\gamma_{10} = 0.4$ in the second test case. We intentionally consider the lower fleet size of the ride-hailing service (capacity 1) to make its level of service poorer than the one achieved with the initialized waiting time and travel time. The share of each mode is shown in Figure~\ref{fig:test_case2}. As expected, the mode share of the ride-hailing service quickly decreases in the first 5 iterations. According to initial conditions, the ride-hailing service is attractive but its demand rapidly decreases as passengers learn that the service is not able to serve the demand. In addition, the mode share of the ridepooling service increases rapidly in the first 5 iterations as it offers a higher level of service than initial conditions due to a high fleet size of 1000 vehicles. The stop criterion value for the last five iterations are reported. For test case 1, the stop criterion value for the last 5 iterations are 0.0031, 0.0029, 0.0018, 0.0027 and 0.0027. For test case 2, the stop criterion value for the last 5 iterations are 0.0045, 0.0025, 0.0027, 0.0028, and 0.0018. \par

Note that even with a fleet size of 100 vehicles in the second test case, the ride-hailing service could attain the larger share than both high capacity MoD services. There are two reasons for this to happen: i) The demand data of Manhattan contains many short trips. The passenger is likely to choose the ride-hailing service over high capacity services for short trips because even the discount cannot compensate for the lower level of service in high capacity MoD; ii) As discussed earlier in Section~\ref{sec: formulation}, we assume that unsatisfied MoD demand (when the waiting time is larger than the threshold) shifts to the public transit and the penalty on the utility of MoD service is applied in the next iteration. It appears that even though the service rate for ride-hailing service is low, the penalty is not high enough to shift the demand to other high capacity MoD services. \par

Additionally, the public transit share is very high (35\% - 40\%) in both cases, even though the input data corresponds to trips that were taken by taxi in reality. This is a manifestation of not having an alternative specific constant (ASC) for transit in the mode choice models. Therefore, the ASC of the public transit needs to be calibrated to lower its mode share for this specific travel demand considered in this study, which we describe next. \par 

\subsection{Calibrating the Alternative Specific Constant of Transit. }\label{sec: ASC}
In the absence of revealed preference data, calibration of ASCs is difficult. Since ride-hailing service share characteristics of Uber, and ridepooling and micro-transit services are similar to UberPool, we use their respective ASCs as our best guess in the case study. Moreover, MNL estimates of marginal utilities of alternative-specific attributes are used in the case study to ensure consistency of willingness to pay and scale. \par

\begin{figure}[htb]
\centering
\includegraphics[width=0.65\textwidth]{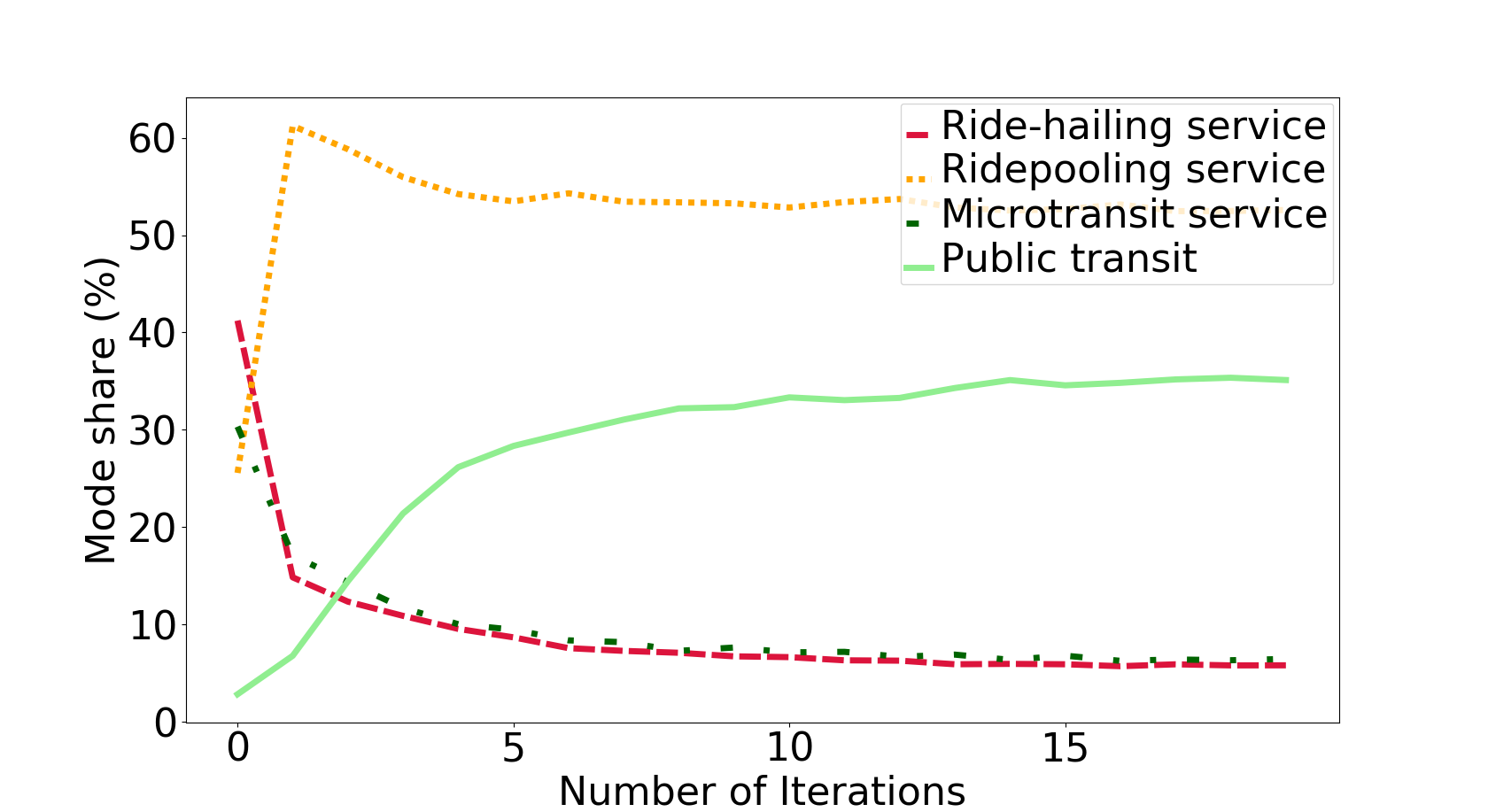}
\caption{\label{fig:test_case2_2} The mode share for test case 2 with the calibrated ASC of transit.}
\end{figure}

If we had the true share of transit for the given demand, then the ASC of transit could have been directly calibrated using the method suggested by \cite{train2009discrete}. Since, the travel demand considered in the study corresponds to trips that actually used the taxi as travel mode, a transit mode share of 0-10\% in our results was endogenously set as an appropriate target. To calibrate the transit ASC of the choice model for this specific population, we use a grid search approach. Since the demand we consider was served by 13,586 active taxis, we ran the inner loop simulation at different values of transit ASC with only transit and a fleet of 13,586 ride-hailing vehicles. The share for public transit for various ASC values are shown in Table~\ref{tab: calibrate}. Transit attains a share of 6.3\% at an ASC value of -3, which appears appropriate for the considered travel demand and thus is used in the remaining experiments. Figure~\ref{fig:test_case2_2} shows the results of running the test case 2 in Section~\ref{sec: equilibrium} with -3 as the ASC for public transit. The shares of the ride-hailing and the ridepooling services become $5.8\%$ and $52.6\%$ which were $20.8\%$ and $19.9\%$ when ASCs were not considered (see Section~\ref{sec: equilibrium}). \par

\begin{table}[]
\centering
\caption{ASC and resulting mode share for public transit. }
\label{tab: calibrate}
\begin{tabular}{@{}cc@{}}
\toprule
ASC  & Mode share (\%) \\ \midrule
-1.0 & 32.9            \\
-1.5 & 23.2            \\
-2.0 & 15.5            \\
-2.5 & 10.2            \\
-3.0 & 6.3             \\ \bottomrule
\end{tabular}
\end{table}

\subsection{Algorithm Performance.}\label{sec: alg_performance}
In this section, we illustrate the application of the proposed BO framework to optimize the supply-side parameters and evaluate its performance by: (1) Use a smaller test example (10\% of the real travel demand) to compare BO solution with the brute-force solution; (2) Compare the performance of BO and random search method \citep{bergstra2012random} using real-scale travel demand. \par

Optimizing the supply-side parameters for an MoD system is analytically intractable due to the complexity of the supply-demand interactions, since the fleet assignment problem is a function of the demand and vice versa. Therefore, it is difficult to find an optimal solution even for relatively small test cases. Instead, we use the following approach to approximate the optimal solution via a grid search of relatively high resolution and compare that solution to that of BO. First we decrease the travel demand to 10\%, and change the maximum value for $n_1$, $n_4$ and $n_{10}$ to be 800, 300 and 150 respectively. The discount factors $\gamma_4$ and $\gamma_{10}$ are set to constants (0.1 and 0.2, respectively) to further decrease the search space. In this setting, we enumerate every possible combination of the fleet size in $\{(n_1, n_4, n_{10}) \mid n_1 \in [25, 50, ..., 800], n_4 \in [0, 25, ..., 300], n_{10} \in [0, 25, ..., 150]\}$ and find the solution with the highest profit as the near-optimal solution. On the same test case, we employ BO with three different acquisition functions (UCB, PI and EI) to understand the sensitivity of BO's performance relative to the acquisition functions by comparing the gap between the respective BO solutions and the near-optimal solution found by the full enumeration method. \par

\begin{figure}[htb]
\centering
\includegraphics[width=0.65\textwidth]{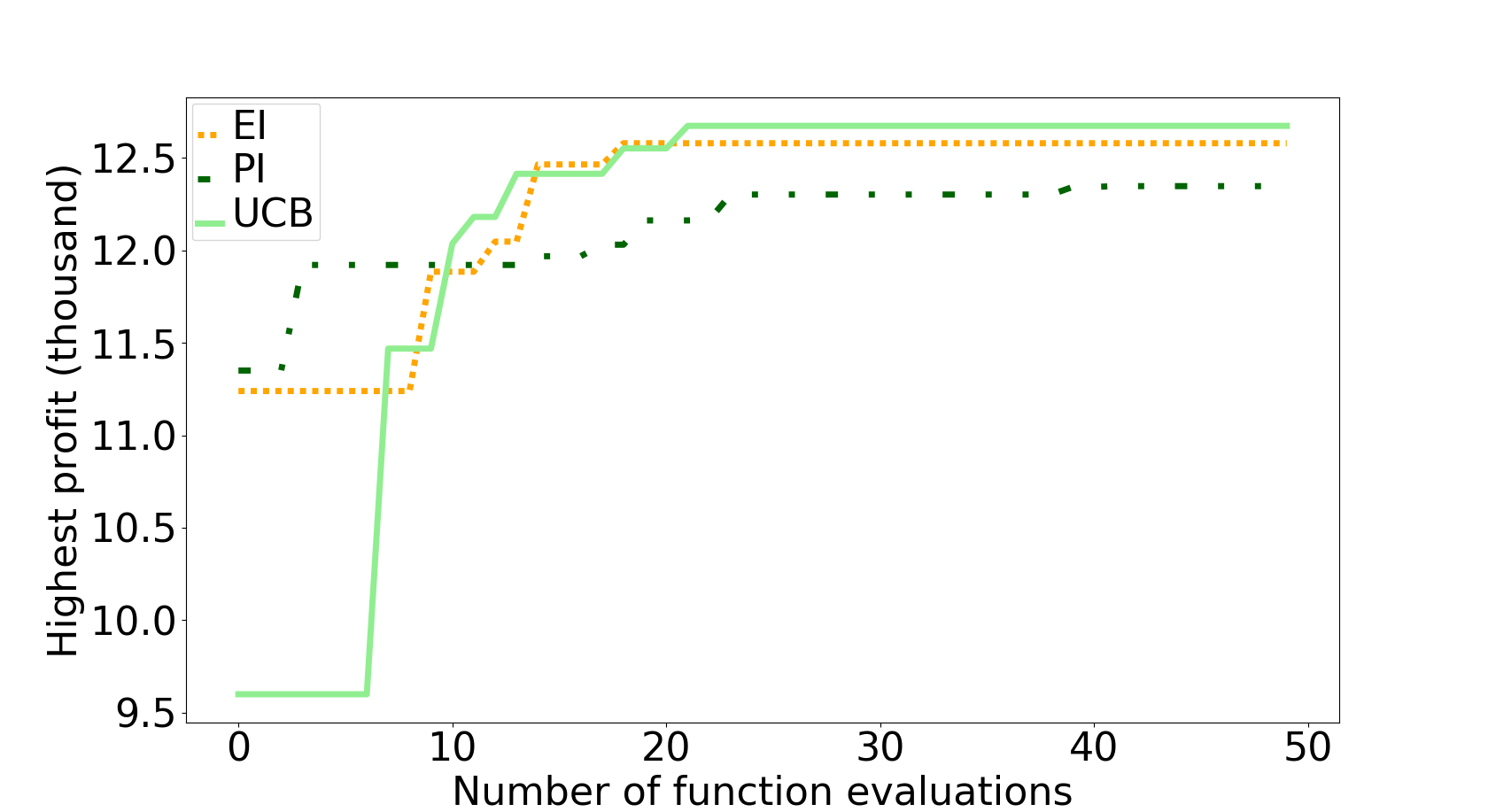}
\caption{\label{fig: acq} The comparison of the  acquisition functions' performance.}
\end{figure}

In the full enumeration method, we evaluate the profit for 2912 parameter settings. The best solution is obtained with $n_1 = 225$, $n_4 = 150$, $n_{10} = 0$, and provides a profit of 13001. Figure 5 shows the variation in the highest profit with the number of function evaluations for BO using three different acquisition functions. It can be seen that PI performs worse than EI and UCB, while EI and UCB perform similarly well in our example. The best solution found by BO is $n_1 = 182$, $n_4 = 85$, $n_{10} = 99$, which gives a profit of 12674. Although the optimal supply-side parameters provided by BO are quite different from the near-optimal solution of the full enumeration method, the profit gap is only about 2.5\%. \par

In the second experiment, we use real-scale travel demand and compare the performance of BO and random search method. In the random search method, we sample the decision variables from a predefined distribution (uniform distribution in our case), evaluate the objective function at those realizations, and report the realization of the decision variable corresponding to optimal objective function value as the optimal solution. For a fair comparison, we keep the same number of function evaluations for BO and the random search method. \par

\begin{figure}[htb]
\centering
\includegraphics[width=0.65\textwidth]{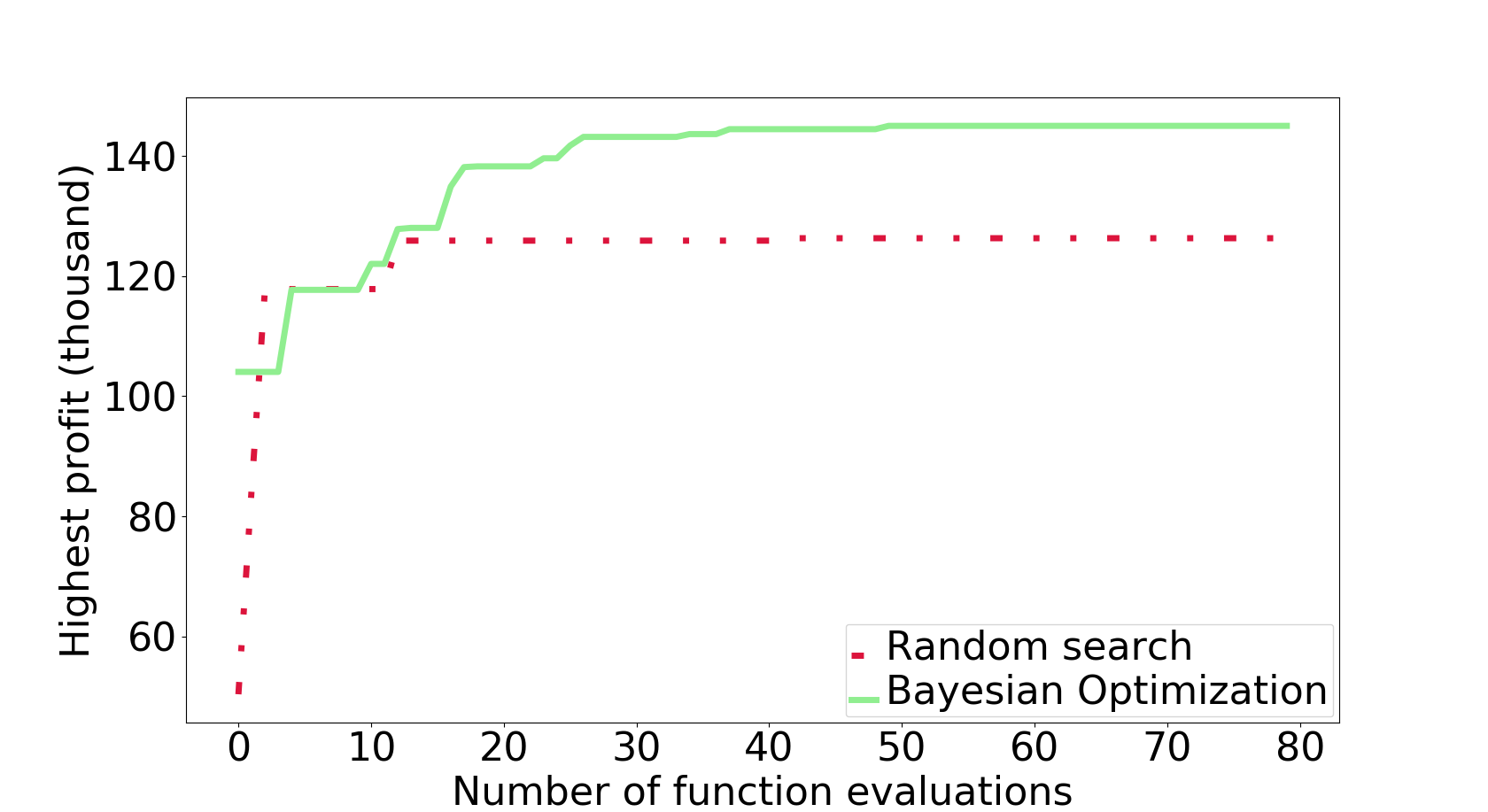}
\caption{\label{fig:alg_performance} The comparison of the algorithms' performance.}
\end{figure}

Figure~\ref{fig:alg_performance} shows the variation in the highest profit with the number of function evaluations in both optimization methods. The highest profit bound by BO and the random search are 145,015 and 126,308 respectively, which represents that BO's solution increases the profit by 15\% compared to the random search's solution. This considerable gap between objective functions shows the advantage of using BO over the random search. We expect this gap to be larger in problems with more decision variables i.e., higher dimensional solution spaces. \par

The BO optimal solution is: $n_1 = 817$, $n_4 = 1539$, $n_{10} = 173$, $\gamma_4 = 0$ and $\gamma_{10} = 0$. It is important to note that the discount factor is 0 for both high capacity MoD services, which seems quite unrealistic because passengers are not likely to use high capacity MoD services if they have to pay the same price as of ride-hailing service. This experimental result implies that discounting the price of high capacity MoD services is not able to attract a level of demand that is high enough to compensate the reduction in profit due to the discounting. In general, the passengers have different perceptions about the discounted fare of the high capacity MoD services, but such latent factors are hard to quantify. The current mode choice model we use assumes that the trade-off between sharing a ride and the cost savings of doing so can be modeled as a linear relationship, and does not consider the fact that travelers perceptions of this trade-off may be more nuanced. \par

\subsection{Scenario Analysis. }\label{sec: scenario}
Modeling passenger's perception of discounting is left for future research but we illustrate implications of including such latent factors in choice models using a scenario-based analysis. We hypothesize that a passenger is less likely to choose high capacity MoD services at low values of the discount factor, even if these services have the same attribute-governed-utility\footnote{The attribute-governed-utility implies the part of utility which depends on observed attributes of the travel mode such as travel time, waiting time and travel cost in our case.} as of the ride-hailing service. We represent this hypothesis by adding a function of the discount factor, which is denoted as $f(\gamma)$, in the utility equation of the high capacity MoD services: \par

\begin{equation}
    f(\gamma) = \min(0, a + b \cdot e^{-c \cdot \gamma})
\end{equation}

\begin{figure}
\begin{minipage}[t]{0.5\linewidth}
\centering
\includegraphics[width=3.6in]{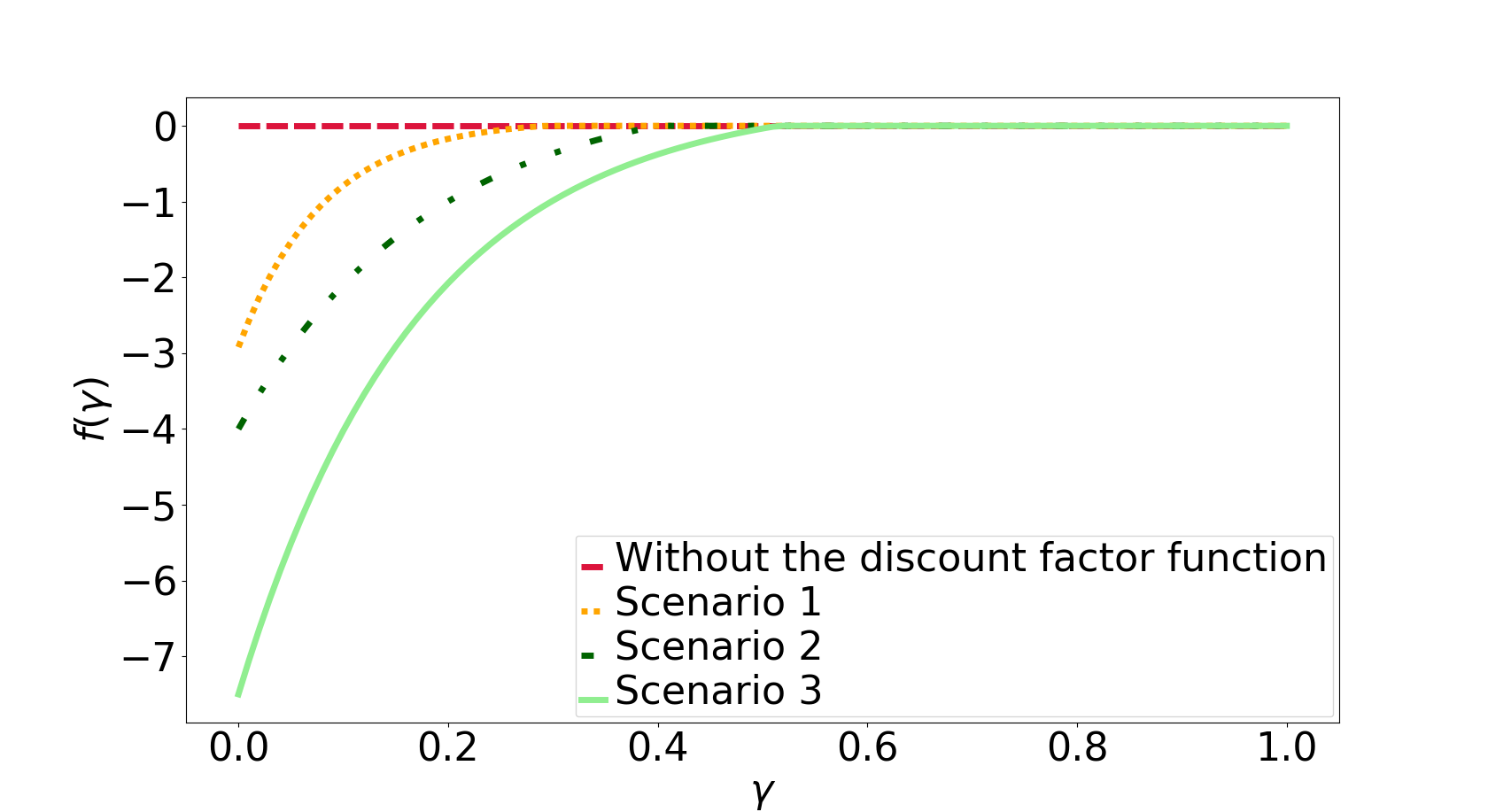}
\caption{\label{fig: gamma_4} $f(\gamma)$ for the capacity 4 service.}
\end{minipage}%
\begin{minipage}[t]{0.5\linewidth}
\centering
\includegraphics[width=3.6in]{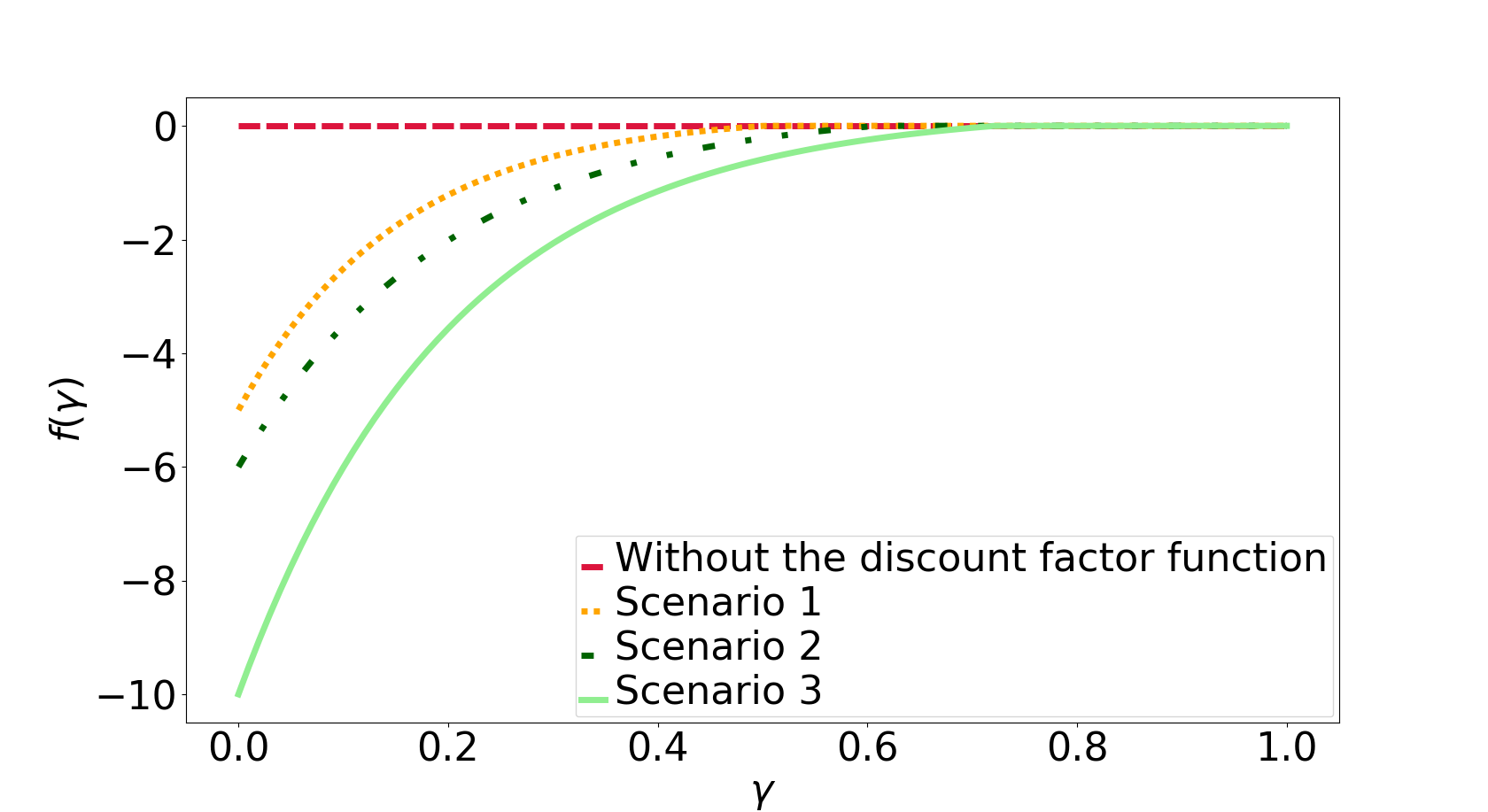}
\caption{$f(\gamma)$ for the capacity 10 service.}
\label{fig: gamma_10}
\end{minipage}
\end{figure}

We call the function of the discount factor that we add to the utility equation \textit{discount factor function}. We optimize MoD operations under three different discount factor functions, which are shown in Table~\ref{tab: discount_f}. The coefficients in $f(\gamma)$ are set such that three scenarios represent low disutility, medium disutility and high disutility for having a low discount factor of high capacity MoD services. Since the passenger is likely to expect a higher discount for the micro-transit service than the ridepooling service, we use different values of coefficients in both functions to mimic these expectations. The plots of discount factor functions for the ridepooling and micro-transit services under three scenarios are shown in Figures~\ref{fig: gamma_4} and~\ref{fig: gamma_10}. \par

\begin{table}[]
\centering
\caption{The discount factor functions for each scenarios. }
\label{tab: discount_f}
\begin{tabular}{@{}ccc@{}}
\toprule
Scenario & \begin{tabular}[c]{@{}c@{}}Discount factor function\\ for the ridepooling service\end{tabular} & \begin{tabular}[c]{@{}c@{}}Discount factor function\\ for the micro-transit service\end{tabular} \\ \midrule
1        & $0.08 - 3\cdot e^{-12.4\cdot \gamma}$                                                                                             & $0.2 - 5.2\cdot e^{-6.5\cdot \gamma}$                                                                                               \\
2        & $0.5 - 4.5\cdot e^{-5.5\cdot \gamma}$                                                                                              & $0.3 - 6.3\cdot e^{-5\cdot \gamma}$                                                                                                \\
3        & $0.4 - 7.9\cdot e^{-5.8\cdot \gamma}$                                                                                             & $0.3 - 10.3\cdot e^{-4.9\cdot \gamma}$                                                                                               \\ \bottomrule
\end{tabular}
\end{table}

\begin{table}[]
\centering
\caption{The solution for each scenario. }
\label{tab: solutions}
\begin{tabular}{@{}ccccccc@{}}
\toprule
Scenario                         & $n_1$ & $n_4$ & $n_{10}$ & $\gamma_4$ & $\gamma_{10}$ & Profit \\ \midrule
With no discount factor function & 817  & 1539  & 173   & 0  & 0   & 145015      \\
1                                & 2826  & 235  & 526   & 0.12 & 0.24  & 122838     \\
2                                & 2801 & 428 & 690  & 0.17 & 0.29  & 119154     \\
3                                & 2900 & 924 & 19  & 0.33 & 0.69  & 111141     \\ \bottomrule
\end{tabular}
\end{table}

The optimal supply-side parameters under three scenarios were found using the BO-based approach and results are summarized in Table~\ref{tab: solutions}. The discount factors of all MoD services are positive after employing discount factor functions, and $\gamma_{10}$ is (as expected) higher than $\gamma_4$. As the disutility of high capacity MoD services at low discount factors increases in scenario 2, the service provider has to offer more discount and increase the fleet size to attract the passengers towards high capacity services, and thus the profit also decreases. Note that the fleet size of micro-transit (capacity 10) is only 19 in scenario 3. This result represents the situation when the disutility of using micro-transit at a low discount factor is extremely high, and thus the service provider finds it hard to make profits by running this service. \par

\begin{table}[]
\centering
\caption{The mode share for each scenario. }
\label{tab: scenario_share}
\begin{tabular}{@{}ccccc@{}}
\toprule
Scenario                                                                    & \begin{tabular}[c]{@{}c@{}}Ride-hailing\\share (\%)\end{tabular} & \begin{tabular}[c]{@{}c@{}}Ridepooling\\ share (\%)\end{tabular} & \begin{tabular}[c]{@{}c@{}}micro-transit\\ share (\%)\end{tabular} & \begin{tabular}[c]{@{}c@{}}Public transit\\ share (\%)\end{tabular} \\ \midrule
\begin{tabular}[c]{@{}c@{}}With no discount \\ factor function\end{tabular} & 23.2                                                                 & 56.1                                                                 & 13.7                                                                  & 7.0                                                                             \\
1                                                                           & 61.0                                                                 & 13.5                                                                 & 19.2                                                                  & 6.3                                                                             \\
2                                                                           & 60.5                                                                 & 14.8                                                                 & 18.6                                                                  & 6.1                                                                             \\
3                                                                           & 62.2                                                                 & 26.4                                                                 & 6.4                                                                   & 5.0                                                                             \\ \bottomrule
\end{tabular}
\end{table}

\begin{figure}[htb]
\centering
\includegraphics[width=0.7\textwidth]{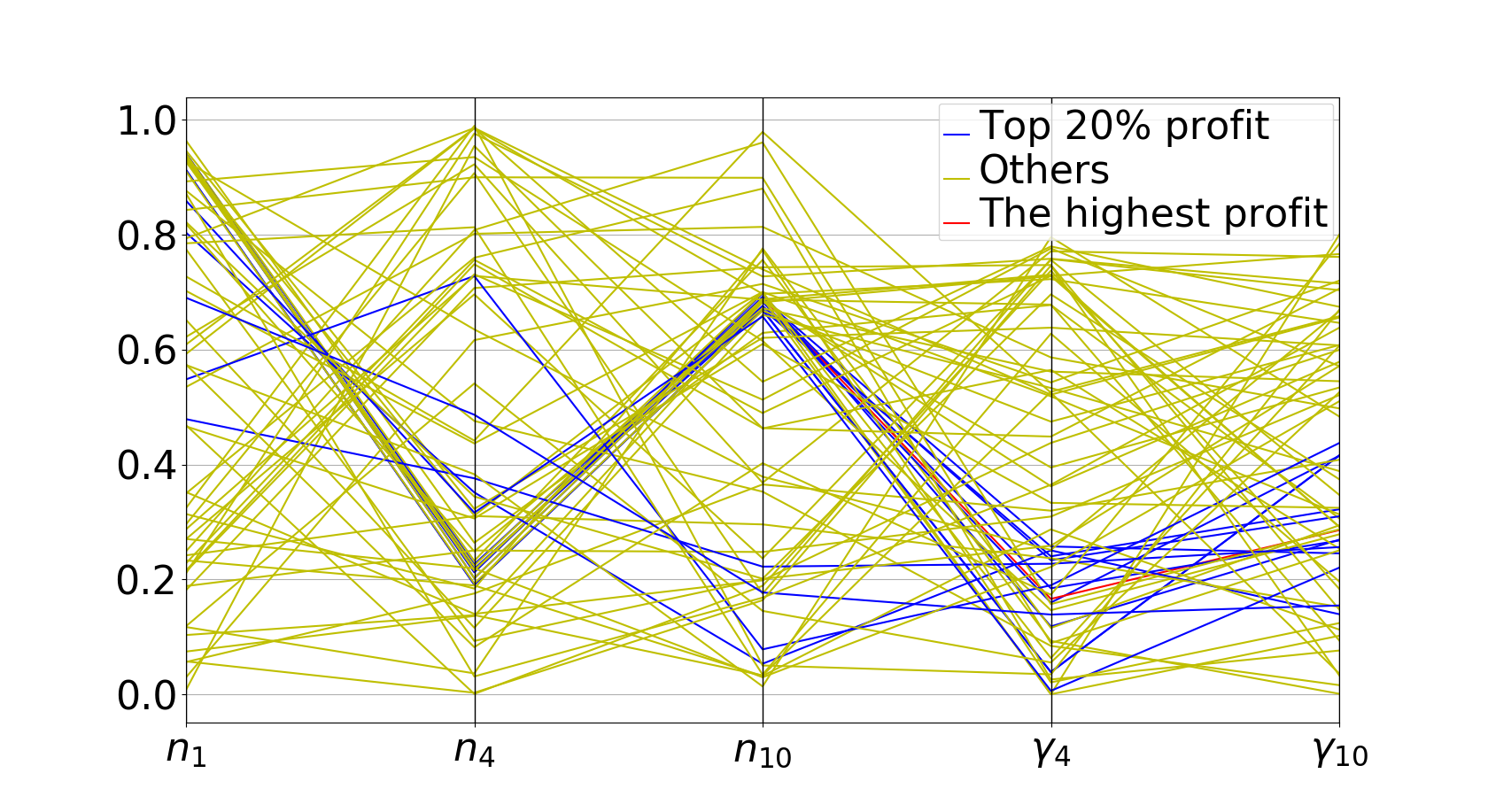}
\caption{\label{fig:scenario2_solutions} Decision variables at various function evaluations in the BO framework (scenario 2).}
\end{figure}

Table~\ref{tab: scenario_share} shows the share of each mode in all scenarios. In a base case scenario when discount factor function is ignored, a large portion of the demand (around 70\%) is served by high capacity MoD services. However, in scenarios with a discount factor function, a large fraction of demand (around 60\%) is served by the ride-hailing service. Therefore, the operating cost of the service provider appears to increase in order to make the same revenue. An increase in the operating cost and a higher discount factor of high capacity MoD services result in a lower profit. Not only the share of micro-transit service decreases in scenario 3, its early demand shifts to the ridepooling service instead. This shift can be attributed to the similar values of $f(\gamma)$ for the ridepooling service in scenario 3 and for the micro-transit service in scenario 2 (see Figure~\ref{fig: gamma_4} and Figure~\ref{fig: gamma_10}). \par

Figure~\ref{fig:scenario2_solutions} shows a parallel coordinates plot of decision variables for which the BO framework evaluates the objective function in process of searching for the optimal values in scenario 2. For visualization convenience, $n_1$, $n_4$ and $n_{10}$ are normalized by dividing with their maximum values (3000, 2000 and 1000 respectively). The red, blue, and yellow lines represent the best, the top $20\%$, and all remaining solutions, respectively. The figure shows that the BO framework explores a broad domain of the solution space while exploiting the solution space with high expected profits more intensively. \par

\begin{figure}
\begin{minipage}[t]{0.5\linewidth}
\centering
\includegraphics[width=3.6in]{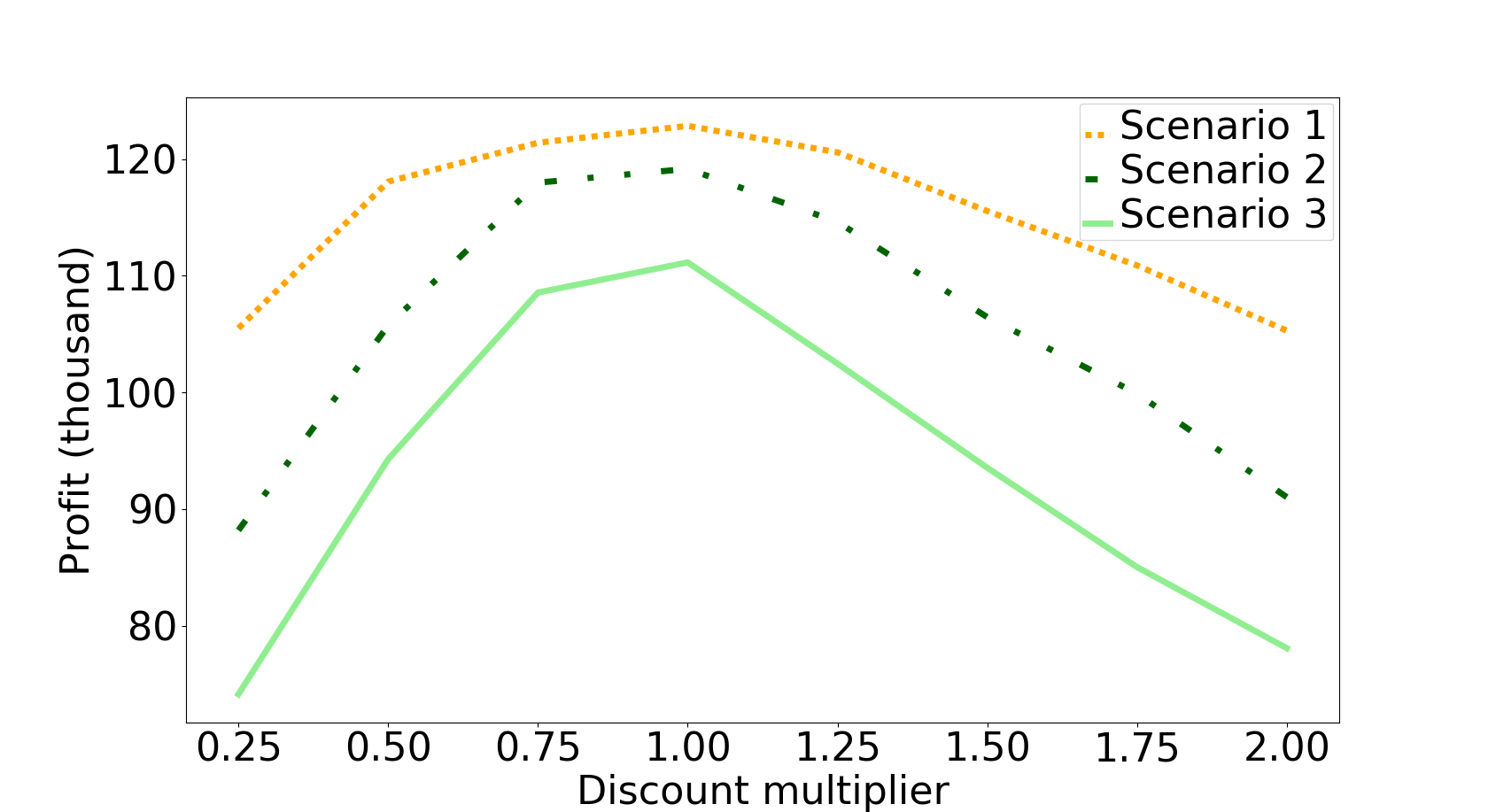}
\caption{\label{fig: profit_discount} The profit for different discount multipliers and different scenarios.}
\end{minipage}%
\begin{minipage}[t]{0.5\linewidth}
\centering
\includegraphics[width=3.6in]{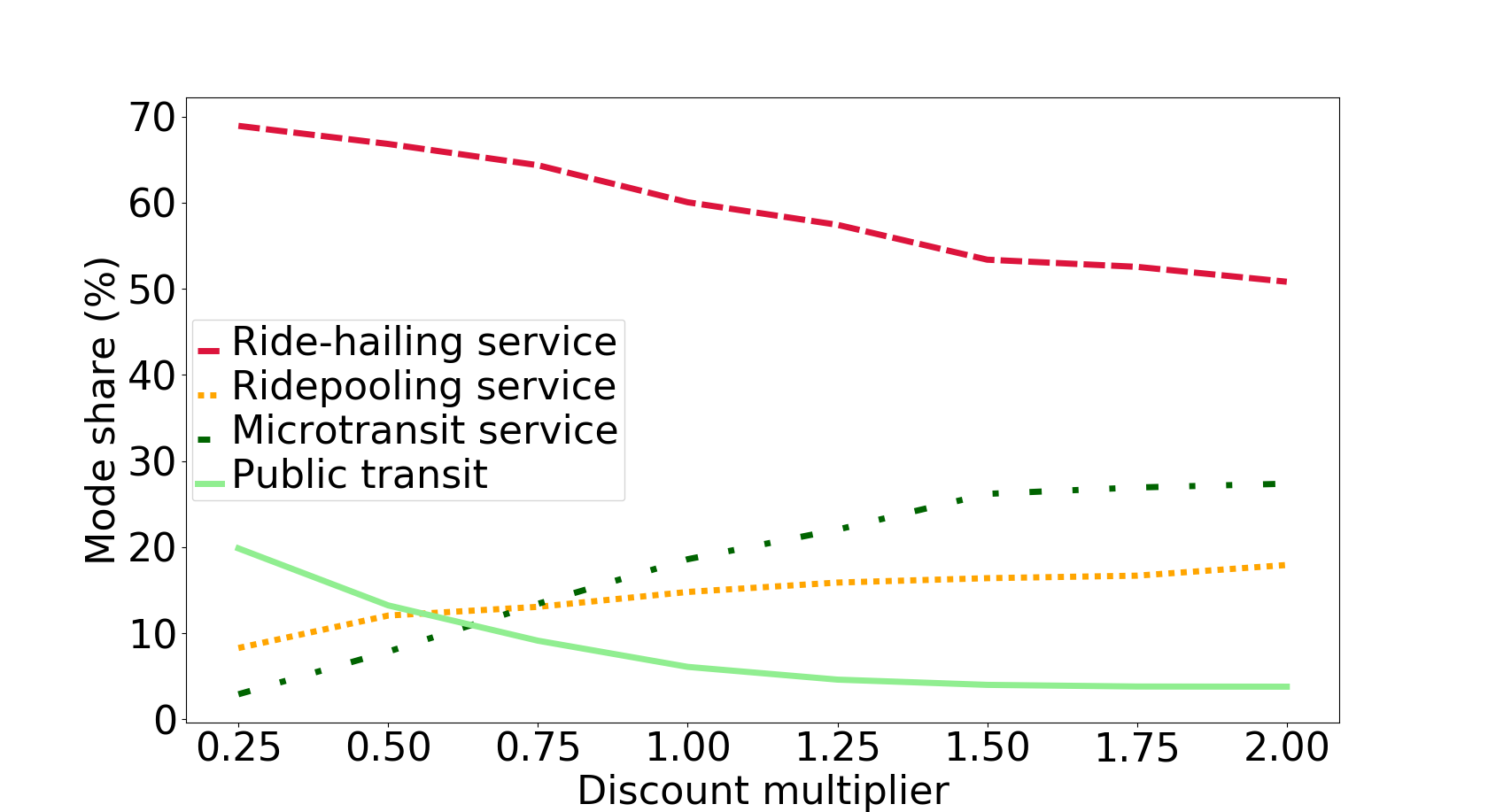}
\caption{The mode share for Scenario 2 with different discount multipliers. }
\label{fig: mode_share_discount}
\end{minipage}
\end{figure}

In the experiments discussed above, we change both the discount factors and the fleet sizes for each MoD service to find the best solution using BO. It is difficult to infer the effects of discount factors on the mode share and the profit of MoD service providers. Therefore, we conduct an additional set of experiments for each scenario mentioned above. In these experiments, the fleet size for each MoD service is kept constant, which is the solution reported in Table~\ref{tab: solutions}. We use different multipliers in $[0.25, 0.5, 0.75, 1.00, 1.25, 1.50, 1.75, 2.00]$ to increase or decrease the discount factors reported in Table~\ref{tab: solutions} for each scenario. For example, if the multiplier we used for Scenario 2 is 0.75, then $\gamma_4 = 0.75 \cdot 0.17 = 0.1275$. Figure~\ref{fig: profit_discount} shows the profit for three scenarios with different discount multipliers. The profit first increases and then quickly decreases as the discount factors increase. The highest profit is given by the discount factors found by BO. Figure~\ref{fig: mode_share_discount} shows the mode share for Scenario 2 with different discount factor multipliers. We should note that: (i) As the discount increases for MoD services, the mode share of public transit service keeps decreasing as it gradually loses the advantage of low cost; (ii) The mode share of ridepooling and micro-transit service increases as the discount factor increases, but the mode share of micro-transit service increases faster. The reason is that the base discount factor for micro-transit service is larger than ridepooling service, and thus the discount factor increases more as the discount multiplier increases. \par 

\subsection{An Application: Per-ride Tax}
In this section, we illustrate how the proposed framework can be used to quantify the impact of system-level policy interventions. We consider a scenario where the government plans to impose a per-ride tax on only the ride-hailing (capacity 1) MoD service. This type of tax has already been considered in California \citep{arstechnica.com}. \par 

Consider a government policy to charge the MoD service provider \$2 on every ride-hailing trip. In the subsequent analysis, we assume that this cost is absorbed by the MoD provider with no increase to the passenger fare, but similar experiments where the cost is borne completely or partially by the traveler can also be considered. Using the discount factor function in scenario 1, we run the BO framework to optimize the supply-side parameters under this policy intervention. The optimal supply-side parameters, vehicle miles traveled (VMT) of MoD services, the ratio of the passenger miles traveled (PMT) and VMT, and the mode shares under tax and no-tax scenarios are shown in Table~\ref{tab: tax}. Imposing the tax decreases the VMT of MoD services by $10.5\%$ and decreases the profit by $30.5\%$. \par

In this tax scenario, running ride-hailing service is not as profitable as before. However, if the fleet size of the ride-hailing service is lowered, its demand is likely to shift to other travel modes. If the MoD service provider wants to induce this shift to move towards the other MoD modes and not public transit, the service provider needs to offer a higher discount for the shared services and thus the profit may further decrease. The optimal solutions of tax and no-tax scenario validate this hypothesis. A decrease in VMT of MoD services in tax scenario can also be attributed to a shift of demand from ride-hailing services to high capacity MoD services. In fact, the increase in the share of public transit by $1.6\%$ also contributes to lower VMT of MoD services. The tax policy increases the revenue of public transit agency by \$14340 which can further be recirculated for improvement in the level of service of the local public transportation to even further increase the transit ridership. One major benefit of the proposed framework is the ability to perform such scenario analysis and gain insights into the impacts of certain policy interventions.\par

\begin{table}[]
\centering
\caption{Comparison between the case with no tax and with tax.}
\label{tab: tax}
\begin{tabular}{@{}ccccccc@{}}
\toprule
    & $n_1$    & $n_4$      & $n_{10}$                                                             &  $\gamma_4$                                                             & $\gamma_{10}$                                                             & \textbf{Profit}                                                              \\ \midrule
No tax  & 2826    & 235       & 526                                                               & 0.12                                                             & 0.24                                                              & 145015                                                              \\
\$2 tax & 1388    & 1668      & 31                                                                & 0.15                                                             & 0.67                                                              & 100748                                                              \\ \midrule
        & \textbf{VMT}     & $\frac{\textbf{PMT}}{\textbf{VMT}}$  & \begin{tabular}[c]{@{}c@{}}\textbf{Ride-hailing}\\ \textbf{share (\%)}\end{tabular} & \begin{tabular}[c]{@{}c@{}}\textbf{Ridepooling}\\ 
        \textbf{share (\%)}\end{tabular} & \begin{tabular}[c]{@{}c@{}}\textbf{micro-transit}\\ \textbf{share (\%)}\end{tabular} & \begin{tabular}[c]{@{}c@{}}\textbf{Public transit}\\ \textbf{share (\%)}\end{tabular} \\ \midrule
No tax  & 33835.7 & 1.11      & 61.0                                                              & 13.5                                                             & 19.2                                                              & 6.3                                                                 \\
\$2 tax & 30288.3 & 1.17      & 36.4                                                              & 50.6                                                             & 5.1                                                               & 7.9                                                                 \\ \bottomrule
\end{tabular}
\end{table}

\section{Conclusion}\label{sec: conclusion}
This paper addresses two critical issues in the design of a Mobility-on-Demand (MoD) system, namely modeling MoD services under endogenous demand and optimization of supply-side parameters (e.g., fleet size and fare). We propose a unified framework to design MoD systems that not only allows interaction of MoD services with other competing travel modes, but also optimizes supply-side parameters using Bayesian optimization (BO), a global optimization technique.\par

The proposed framework is calibrated using taxi-demand data from Manhattan County in New York City and a stated preference survey. Demand is served by three types of MoD services (ride-hailing, ridepooling, and micro-transit of capacities 1,4, and 10) and public transit, and user preference for each travel mode is predicted using a choice model. The mode choice model was estimated using stated preference data collected in New York City. We conducted numerical experiments to show the numerical convergence of the integrated supply-demand system and show the application of the BO-based approach in optimizing the supply-side parameters. The BO-based approach outperforms the traditional random search optimization by sequentially taking steps in the direction of potential function improvement. The optimization problem maximizes the profit of MoD service provider in the case study, but the framework is more general and can be used to optimize various objective functions such as consumer surplus, vehicle miles traveled, as well as combinations of these metrics. \par 

This proposed framework provides a potential toolbox for policymakers and planning agencies to evaluate the impact of MoD services on the urban transportation system and vice versa. For instance, we assess the system-level impact of a policy intervention where the government imposes a tax on ride-hailing (capacity 1) trips without allowing the service provider to increase the fare. This policy results in an increase in transit ridership, a decrease in vehicle miles traveled of MoD services, and a reduction in the profit of the MoD service provider due to a need of higher discount factors of high capacity MoD services. Whereas the government can evaluate these policy implications qualitatively, the proposed framework provides a method to properly quantify the consequences of the policy intervention for different stakeholders (government, passenger, MoD service operator). Other such examples include quantifying the benefits for transit agencies if they plan to collaborate with MoD services to better serve first and last mile trips. \par

We now highlight future research avenues to strengthen the proposed framework. The integrated supply-demand system can be viewed as a fixed point problem. Whereas we illustrate the existence of its equilibrium in numerical experiments, an analytical proof or guarantees can increase its credibility and versatility. In addition, we used MNL to predict travel mode preferences because it integrates simply and well with the MoD simulation due to its closed-form choice probability expression. However, we have shown that optimization of supply-side parameters is very sensitive to the parameters of the mode choice model. More complex random parameter (mixed logit) models can be used to possibly mitigate this sensitivity concern but those models require another layer of simulation in the prediction of choice probabilities, which can lead to numerical convergence issues. Thus, developing robust mode choice models with closed-form choice probability expressions that more closely mimic the passengers' decision-making mechanism is an essential avenue for future travel behavior research. In particular, mode choice models for a multimodal system with MoD services are required to go beyond just the level of service attributes (e.g., waiting time, travel time, and travel cost), and various perception-based latent factors (e.g., notion of reliability, variation in disutility of sharing rides with varying number of passengers, perception of discount) need to be incorporated. In addition, we use a simple transit route planner model in this study to estimate transit travel times and do not consider a full-blown transit assignment model. The framework can, however, be extended to include a more advanced transit routing model (e.g. hyperpath-based routing) or a full-blown transit assignment model. Including a transit assignment model will increase the complexity of the overall system, since there will be more dependencies between the supply and the demand (e.g. transit travel-times will also depend on demand and vice versa). \par

\section*{Acknowledgements}
The authors are thankful to the National Science Foundation Award No. CMMI-1462289 for financially supporting this research. We are also thankful to two anonymous referees for their constructive comments.





\bibliographystyle{elsarticle-harv}

\bibliography{sample}

\end{document}